\documentclass[aps,prb,twocolumn,showpacs,floatfix,superscriptaddress]{revtex4-2}

\usepackage{graphicx,color}
\usepackage{amsfonts}
\usepackage[figuresright]{rotating}  
\usepackage{amssymb}
\usepackage{amsmath}
\usepackage{ragged2e}
\usepackage{blindtext}

\usepackage{bbold}
\usepackage{wrapfig,lipsum,booktabs}
\usepackage{mathtools}
\usepackage{psfrag}
\usepackage{subfigure}
\usepackage{multirow}
\usepackage{tabularx}
\usepackage{textcomp}
\usepackage{bm}
\usepackage{hyperref}
\usepackage{relsize}
\hypersetup{
 pdfnewwindow=true, colorlinks=true,
 linkcolor=blue, anchorcolor=blue,
 citecolor=blue, filecolor=blue,
 menucolor=blue, urlcolor=blue}

\def\nn{\nonumber}

\def\beq{\begin{eqnarray}}
\def\eeq{\end{eqnarray}}

\renewcommand{\v}[1]{\ensuremath{\mathbf{#1}}} 
 
\let\baraccent=\= 
\renewcommand{\=}[1]{\stackrel{#1}{=}} 

\makeatletter

\begin{document}

\title{Energy relaxation dynamics in a nodal-line semimetal}

\author{Benjamin\ \surname{M. Fregoso}}
\affiliation{Department of Physics, Kent State University, Kent, Ohio 44242, USA}

\author{Madhab\ \surname{Neupane}}
\affiliation{Department of Physics, University of Central Florida, Orlando, Florida 32816, USA} 

\author{Anup\ \surname{ Pradhan Sakhya}}
\affiliation{Department of Physics, University of Central Florida, Orlando, Florida 32816, USA}

\begin{abstract}
We study the temperature relaxation dynamics of nodal-line semimetals after a sudden excitation in the presence of acoustic and optical phonon modes. We find that the nodal line constrains the electron momenta in scattering processes and,  as a result, the temperature relaxation due to acoustic phonons is exponential as a function of time. However, depending on initial conditions, other functional forms are possible. In typical pump-probe experiments, the temperature relaxation is linear due to acoustic phonons with rates that vary as $\sim n^{1/2}$ with density. The temperature relaxation due to optical phonons is also linear with rates $\sim n^{-1/2}$ or $\sim n$. 
\end{abstract}

\maketitle

\section{Introduction} 
Energy dissipation via phonon scattering is the main pathway to thermalization in electronic systems~\cite{Gantmakher1987,Sze2021,Allen1987,Wellstood1994,Massicotte2021}.  Energy relaxation in metals has been extensively studied in the context of pump-probe experiments where electrons are excited via short laser pulses. After the initial excitation, electrons lose energy to the lattice degrees of freedom and eventually reach equilibrium. 

In a typical pump-probe experiment, the temperature initially falls rapidly due to optical phonon emission. Below certain crossover temperature, optical phonon emission becomes less efficient and a slower decay associated with acoustic phonons sets in. While this scenario is understood, the effects of the Fermi surface (FS) topology on the relaxation dynamics are just beginning to be explored. For example, it is known that at low temperature, the so-called phonon cooling power scales as $\mathcal{P}\propto T_e^{\kappa}- T_L^{\kappa}$, with the instantaneous electron temperature $T_e$ and the lattice temperature $T_L$. For a conventional three-dimensional (3D) metal \cite{Gantmakher1987,Wellstood1994}, $\kappa=5$, but for thin films \cite{Sergeev2000,Karvonen2005}, $\kappa=6$. If the phonon phase space is constrained \cite{Hekking2008}, $\kappa= 3$, and for graphene \cite{Kubakaddi2009}, $\kappa=4$.

Of recent interest is the relaxation dynamics of Dirac materials which have zero-energy manifolds embedded in the Brillouin zone \cite{Armitage2018}. For example, graphene or topological insulators are two-dimensional (2D) semimetals with zero-energy points (Dirac points). Near the Dirac point the quasiparticle dispersion is linear in momentum. As a consequence of its small FS, graphene exhibits strong suppression of electron-phonon scattering below the Bloch-Gruneisen temperature \cite{Stormer1990,Efetov2010} and novel disorder-mediated electron-phonon scattering \cite{Tse2009,Song2012,Graham2013}.  Graphene also exhibits $T_e\sim 1/\sqrt{t}$ temperature relaxation as a function of time due to acoustic phonon relaxation \cite{Bistritzer2009,Viljas2010,Wang2012}.

Weyl (or Dirac) semimetals (SMs) are different Dirac materials in that they are 3D SM with zero-energy (Weyl) points. Near Weyl points, the FS is a sphere instead of a circle. They exhibit asymmetric Fano line shapes \cite{Coulter2019} and strong electron-phonon coupling constant \cite{Osterhoudt2021}. The temperature relaxation in Weyl SMs varies as $T_e \sim 1/\sqrt[3]{t}$ (at long times) due to acoustic-phonon scattering \cite{Lundgren2015}. 

The discovery of nodal-line semimetals (NLSM) \cite{Burkov2011,Kim2015} added another member to the family of Dirac materials. NLSMs are 3D semimetals with zero-energy \textit{lines} in momentum space. NLSMs can also be classified by topological invariants\cite{Fang2015,Kim2015,Fang2016} and were demonstrated experimentally in PbTaSe$_2$~\cite{Bian2016}, PtSn$_4$~\cite{Wu2016}, and ZrSiS~\cite{Schoop2016,Neupane2016} with other materials being investigated \cite{Yang2018,Klemenz2019}. NLSMs exhibit unusual magnetoresistance \cite{Ali2016}, strong light-matter interactions \cite{Gatti2020}, and correlation-induced reduction of free carrier Drude weight \cite{Shao2020}. They are predicted to exhibit a quasi-topological electromagnetic response \cite{Ramamurthy2017}, enhanced excitonic instability \cite{Rudenko2018}, poor coulomb screening \cite{Huh2016}, and diverging mobility \cite{Syzranov2017}. 

\begin{table}[b]
\caption{Temperature scaling of the cooling power $\mathcal{P}_a$ and heat capacity $\mathcal{C}$ for NLSM, Graphene, and Weyl SMs. $\mathcal{C}$ is sensitive to the volume of the FS, whereas the $\mathcal{P}_a$ to the topology of the FS.  NLSMs have small $\mathcal{P}_a$ and small $\mathcal{C}$, which result in exponential temperature relaxation. Graphene and Weyl SMs, on the other hand, have slower $1/\sqrt{t}$ and $1/\sqrt[3]{t}$ temperature relaxations, [see Eq.(\ref{eq:dTe_dt})]. }
\begin{center}
\begin{tabular}{l c c c} 
 \hline \hline
    	~~~~~~~~~~~~~~~~~~~~~~~~~~ 	& ~~~~NLSM~~~        &  ~~~Graphene~~~      &  ~~~Weyl~~~       \\ [0.5ex] 
 \hline 
 Spatial dimension          			    & 3           &  2             & 3            \\ 
 Nodal dimension                      & 1           &  0             & 0             \\  
 $\mathcal{C}\sim T_e^{c}$, ~~$c=$   	& 2           &  2             & 3             \\  
 $\mathcal{P}_a \sim T_e^{p}$, $p=$   & 3           &  5             & 7             \\  
 $\mathcal{P}_a/\mathcal{C} \sim T_e^{p-c}$~~  & 1  & ~~3$^{a}$     & ~~4$^{b}$          \\  
 \hline \hline
 \end{tabular}
\label{table:dim_nlsm_intro}
\end{center}
\begin{FlushLeft}
$^{a}$Reference~\cite{Bistritzer2009}.\\
$^{b}$Reference~\cite{Lundgren2015}.
\end{FlushLeft}
\label{table:intro_summary}
\end{table}

\begin{table*}
\caption{Electron temperature as a function of time in NLSMs, graphene, and Weyl SMs in various limits. $T_e$ is the electron temperature, $\omega_0$ is the optical phonon energy, $\mu$ is the chemical potential, $T_L$ is the lattice temperature,  exp indicates exponential, lin indicates linear, and inv log indicates inverse log. By low density we mean $\mu\ll k_B T_e$ and by high density $\mu \gg k_B T_e$. Only the tails of the relaxation functions are shown. The reference indicates the relevant equation in the text. For clarity we omit $\hbar$ and $k_B$. }
\begin{center}
\begin{tabular}{l c c c} 
\hline \hline
          		& ~~~~~~~~~~~~~NLSM~~~~~~~~~~~~~ &  ~~~~~~~~~~~~~Weyl SM~~~~~~~~~~~~~ & ~~~~~~~~~~~~~Graphene~~~~~~~~~~~~~  \\ [0.5ex] 
\hline
 Spatial dimension 								 & 3               &  3         &  2         \\ 
 Nodal dimension   								 & 1               &  0         &  0         \\ 
\multicolumn{4}{c}{Optical phonon relaxation} \\
Low density: $\omega_{0} \gg T_e \gg  T_L,\mu$ ~~~~~~         & ~~~inv log, ~~~(\ref{eq:op_ph_Te_relax_neutral})   
&  ~~~inv log, ~~~(\ref{eq:relax_won1})      & ~~~inv log, ~~~(\ref{eq:op_ph_Te_relax_neutral})  \\
Low density: $T_e \gg \omega_{0},  T_L,\mu$         & ~~~exp, ~~~~~~~(\ref{eq:Pon2})      
&  ~~~$1/t$, ~~~~~~~(\ref{eq:relax_won2})    & ~~~exp, ~~~~~~~(\ref{eq:Pon2})       \\
High density: $\mu,\omega_{0} \gg  T_e \gg  T_L$    & ~~~inv log,  ~~~(\ref{eq:nlsm_op_doped1})   
&  ~~~~inv log, ~~(\ref{eq:relax_wod1})   & ~~~inv log, ~~~(\ref{eq:nlsm_op_doped1})           \\
High density: $\mu \gg  T_e \gg \omega_{0},  T_L$   & ~~~lin,  ~~~~~~~~(\ref{eq:relax_nlsm_over_case2})      
&  ~~~~lin, ~~~~~~~(\ref{eq:relax_wod2})    & ~~~lin,~~~~~~~~~(\ref{eq:relax_nlsm_over_case2}) \\
\multicolumn{4}{c}{Acoustic phonon relaxation} \\
Low density: $T_e \gg T_L,\mu$              & ~~~exp, ~~~~~~~(\ref{eq:ph_relax_neutral})  
& ~~~~~~~~~~$1/\sqrt[3]{t}$  ~~~~(\ref{eq:weyl_acoustic_neutral}) \cite{Lundgren2015} 
& ~~~~~~~~~$1/\sqrt{t}$  ~~~ (\ref{eq:graphene_acoustic_neutral}) \cite{Bistritzer2009}  \\ 
Low density: $T_e \gtrsim T_L,(\mu=0)$      & ~~~exp, ~~~~~~~(\ref{eq:ph_relax_neutral})      
& ~~~~~~~~~exp, ~~~~~~(\ref{eq:weyl_acoustic_neutral})     \cite{Lundgren2015}          
&  ~~~~~~~~~exp ~~~~~~(\ref{eq:graphene_acoustic_neutral}) \cite{Bistritzer2009} \\ 
High density: $\mu\gg T_e \gg T_L$          & ~~~lin, ~~~~~~~~(\ref{eq:doped_rel})      
& ~~~~~~~~~lin, ~~~~~~~(\ref{eq:relax_wad})  \cite{Lundgren2015}                         
& ~~~~~~~~~lin, ~~~~~~(\ref{eq:relax_gad})   \cite{Bistritzer2009}                \\  
High density: $\mu\gg T_e \gtrsim T_L$      & ~~~exp, ~~~~~~~(\ref{eq:doped_rel})          
& ~~~~~~~~~exp, ~~~~~~(\ref{eq:relax_wad})  \cite{Lundgren2015}                   
&  ~~~~~~~~~exp, ~~~~~(\ref{eq:relax_gad})  \cite{Bistritzer2009}\\  
\hline \hline
 \end{tabular}
\label{table:summary}
\end{center}
\end{table*}

In this paper, we investigate the temperature relaxation dynamics of NLSMs and, in particular, the role of the nodal line. We find that despite the complex FS topology of NLSMs the relaxation due to acoustic phonons is exponential, just as in typical 3D metals. In a NLSM, phonon scattering events constrain as the electron's initial and final momenta to be close to the nodal line. This has two important consequences: (a) at high temperatures the acoustic phonon cooling power is much lower than Weyl SMs or graphene (see Fig.~\ref{fig:ZrSiS-TaAs-graphene_all}), and (b) at low temperatures it gives an exponential temperature relaxation, which is different from a power-law relaxation obtained in Weyl SMs and graphene. To understand this, we write the relaxation equation as
\begin{align}
\frac{d T_e}{dt} \sim - \frac{\mathcal{P}_a}{\mathcal{C}}\sim -T_e^{p-c},
\label{eq:dTe_dt}
\end{align}
where $\mathcal{C} = d\mathcal{E}/d T_e$ is the heat capacity and $\mathcal{P}_a$ the phonon cooling power. As can be seen from Table \ref{table:intro_summary}, $p-c=1$ for NLSMs, $p-c=3$ for graphene, and $p-c=4$ for Weyl SMs.

In a typical pump-probe experiment with an initial temperature $T_0 \sim 100$ meV, Fermi level $\epsilon_F \sim 300$ meV, and optical modes around $\hbar\omega_0 \sim 30$ meV, the predicted relaxation is linear in both the acoustic and optical phonon regimes (see Table~\ref{table:summary}). The density dependence is different in each case (see Sec.~\ref{sec:optica_ph}  and \ref{acustic_ph_cooling_power}).  In fact, linear temperature relaxation is  common to all Dirac materials in this regime. To obtain these results we use a simple two-band model consisting of a ring (nodal line) in momentum space \cite{Kim2015}. Our main assumptions are: (a) the electrons are in equilibrium among themselves at temperature $T_e$, (b) the phonons are in equilibrium among themselves at a fixed temperature $T_L$, and (c) the temperature is higher than the Bloch-Gruneisen temperature. 

The paper is organized as follows. We first consider optical and acoustic phonon branches separately obtaining analytic solutions in various limits  (Secs.~\ref{sec:ph_cooling_power}, \ref{sec:optica_ph} and \ref{acustic_ph_cooling_power}). The solutions are then used to give expressions for the crossover temperature between the optical- and acoustic-phonon regimes in Sec.~\ref{sec:crossoverT}. In Sec.~\ref{sec:numerical_sol} we give numerical solutions including both phonon branches. A simple scaling arguments is given in Sec.~\ref{sec:scaling_arg} to intuitively understand temperature relaxation due to acoustic phonons and we conclude in Sec.~\ref{sec:conclusions}. The appendixes include some calculation details.

\section{Phonon cooling power} 
\label{sec:ph_cooling_power}
A simple mathematical model of the electron's energy loss is 
\begin{align}
\frac{d\mathcal{E}}{dt} = - \mathcal{P},
\label{eq:Ptot_gen}
\end{align}
where $\mathcal{E}(t)$ is the energy of the ensemble assumed to depend on temperature and chemical potential $\mathcal{E}(t)=\mathcal{E}(\mu(t),T_e(t))$, and $\mathcal{P}$ is the rate of energy transfer from the electrons to the lattice, i.e., cooling power \cite{Gantmakher1987}
\begin{align}
\mathcal{E} &= \frac{1}{V}\sum_{n\v{k}} \epsilon_{n\v{k}} f_{n\v{k}},
\label{eq:Etot} \\
\mathcal{P} &= \frac{1}{V}\sum_{n\v{k}} \epsilon_{n\v{k}} \frac{d f_{n\v{k}}}{dt}.
\label{eq:Ptot}
\end{align}

\begin{figure}[]
\subfigure{\includegraphics[width=.44\textwidth]{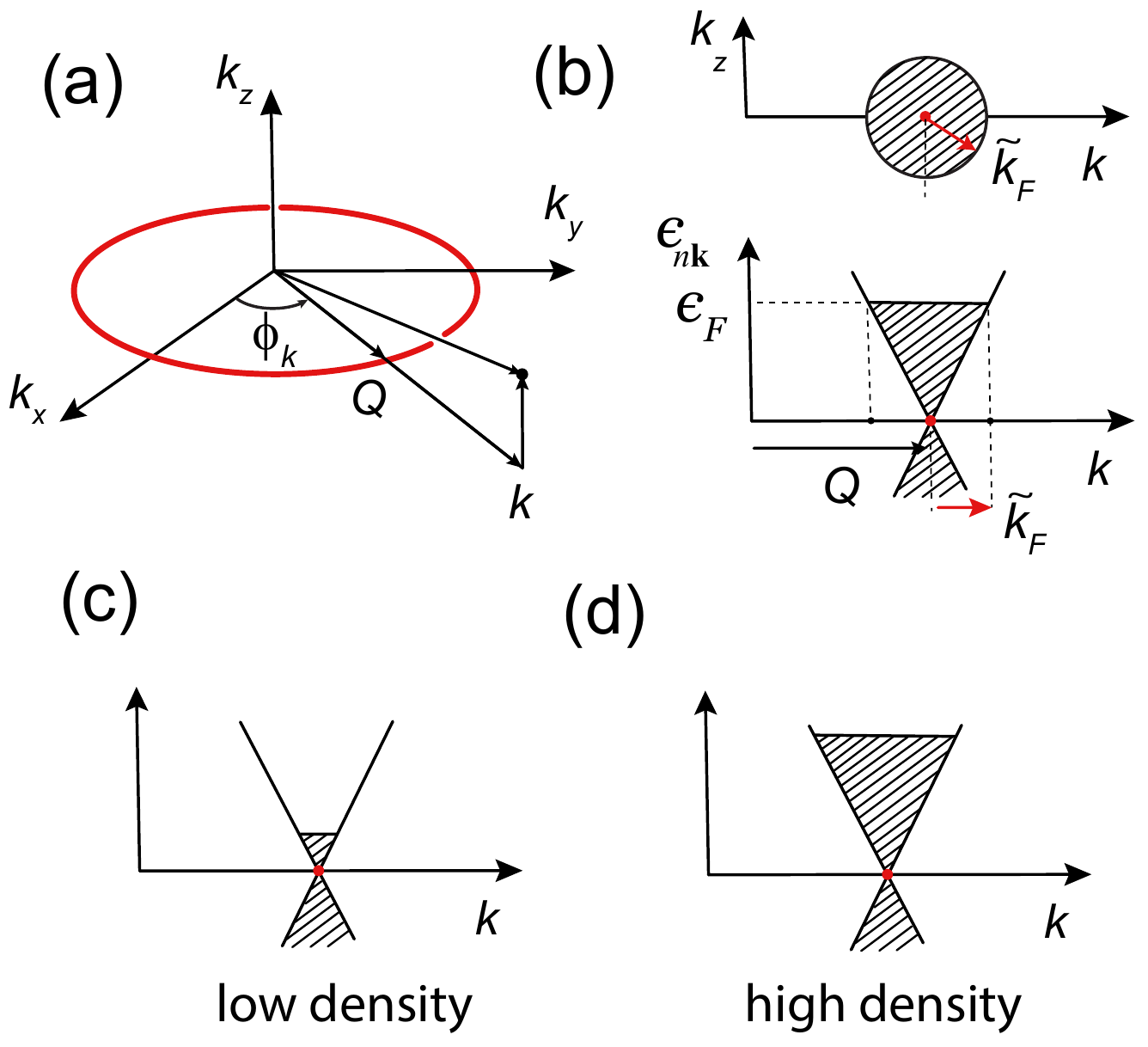}}
\caption{(a) Nodal line (red ring) in a nodal-line semimetal (NLSM) model. (b) Quasiparticles have linear dispersion near points $(k,\phi_{\v{k}},k_z)=(Q,\phi_{\v{k}},0)$, $0\leq \phi_{\v{k}} < 2\pi$. In the plane $k$-$k_z$ the states within radius $\tilde{k}_{F}$ are occupied and the Fermi surface (FS) is a torus. (c) and (d) are two limiting cases considered analytically. The low-density regime is defined by $ 0 \lesssim \mu\ll k_B T_e$ and the high-density regime by $\mu \gg k_B T_e$.}
\label{fig:nodal_ring_geometry}
\end{figure}

Importantly, $\mathcal{P}$ depends on $\mu(t)$ and $T_e(t)$ via the collision integral. $\epsilon_{n\v{k}}$ is the quasiparticle energy. This hydrodynamic approach \cite{Narozhny2021,Lucas2018} assumes that that chemical potential and temperature are well defined at all times; a reasonable assumption if electron-electron interactions thermalize the electron ensemble much faster than phonons.  The collision integral for phonon scattering is
\begin{align}
\frac{d f_{n\v{k}}}{dt} &=  \sum_{m\v{p}}\big[f_{n\v{k}}(1\hspace{-2pt}- \hspace{-2pt} f_{m\v{p}}) W_{n\v{k},m\v{p}} \hspace{-2pt} -\hspace{-2pt}(n\v{k} \leftrightarrow m\v{p})\big],
\label{eq:dfdt} 
\end{align}
with scattering rate
\begin{align}
W_{n\v{k},m\v{p}} &=  \frac{2\pi}{\hbar} \sum_{\v{q}} M_{\v{q}} [(N_{L} \hspace{-2pt}+\hspace{-2pt} 1)\delta_{\v{k},\v{p}+\v{q}} \delta(\epsilon_{n\v{k}}\hspace{-2pt}-\hspace{-2pt}\epsilon_{m\v{p}}\hspace{-2pt}-\hspace{-2pt}\hbar\omega_{\v{q}}) \nn \\
&~~~~~~~~~~~+ N_{L} \delta_{\v{k},\v{p}-\v{q}}\delta(\epsilon_{n\v{k}}\hspace{-2pt}-\hspace{-2pt}\epsilon_{m\v{p}}\hspace{-2pt}+\hspace{-2pt}\hbar\omega_{\v{q}})].
\label{eq:Wmatrix_el}
\end{align}
Here $f_{n\v{k}}\equiv f(\epsilon_{n\v{k}}) = [e^{(\epsilon_{n\v{k}}-\mu)/k_B T_e} + 1]^{-1}$ is the Fermi distribution function, $N_L\equiv N_{L}(\hbar \omega_{\v{q}}) = [e^{\hbar\omega_{\v{q}}/k_B T_L}-1]^{-1}$  the Bose distribution function evaluated at the lattice temperature $T_L$, $T_e$  the electron temperature, $M_{\v{q}}= \hbar^2 D^2 q^2 (1 + s_{nm}\cos\theta)/4\rho V \hbar \omega_{\v{q}}$ the amplitude of electron-phonon scattering, $\v{q}$ the phonon momentum, $\omega_{\v{q}}$ the phonon dispersion relation,  $D$ the deformation potential of acoustic phonons, $\rho$ the ion mass density, $V$ the volume, $\theta$ the angle between $\v{k}$ and $\v{p}$, and $s_{nm}=1(-1)$ intraband (interband) scattering. The quasiparticle dispersion of the NLSM near the ring is
\begin{align}
\epsilon_{n\v{k}} = nv\hbar [k_z^2 + (k- Q)^2]^{1/2},
\label{eq:energ_dispersion}
\end{align}
where $n=1 (-1)$ denotes conduction (valence) band, $v$ is the velocity of nodal quasiparticles, and $Q$ the radius of the nodal ring ( see Fig.~\ref{fig:nodal_ring_geometry}(a)). We consider spinless fermions. From Eqs.(\ref{eq:dfdt}) and (\ref{eq:Wmatrix_el}), we write Eq.(\ref{eq:Ptot}) as \cite{Bistritzer2009,Viljas2010} 
\begin{align}
\mathcal{P} &= \frac{2\pi}{V} \sum_{n\v{k}m\v{p} } \omega_{\v{q}} (f_{n\v{k}}   -  f_{m\v{p}})  (N_L   -   N_e ) M_{\v{q}} \times\nn \\
&~~~~~~~~~~~~~~~~~~~~~~~~~~~~~\delta(\epsilon_{n\v{k}}   -  \epsilon_{m\v{p}}  -  \hbar\omega_{\v{q}}),
\label{eq:P_gen}
\end{align}
(see also appendix \ref{sec:ph_coolingpower}). We defined $\v{q}=\v{p}-\v{k}$. We assume Eq.(\ref{eq:dfdt}) does not vanish even when we have equilibrium distributions of bosons and fermions. The nonvanishing Eq.(\ref{eq:P_gen}) for $T_e \neq T_L$ is consistent with this assumption \cite{Allen1987}. The radius of the nodal line $Q$ is assumed to be the largest momentum scale in the system, i.e., $Q\gg |k_F-Q|,k_B T_e/v\hbar,\omega_0/v$. We also assume phonons with momenta $2Q$ are always thermally excited. This requires a minimum temperature, the Bloch-Gruneisen temperature $T_{BG}$; that is we require $k_B T_e > k_B T_{BG} =\hbar c 2Q$ where $c$ is the speed of sound.

\section{Optical phonon relaxation}
\label{sec:optica_ph}
In this section, we analytically calculate  the temperature relaxation of a NLSM due to optical phonons. We consider a single optical phonon branch with constant energy dispersion $\omega_{\v{q}}= \omega_0$ and constant electron-phonon matrix element $M_{\v{q}}\equiv g^2/V$. From Eq.(\ref{eq:P_gen}), we obtain 
\begin{align}
\frac{d\mathcal{E}}{dt} = - \mathcal{P}_o,
\label{eq:P_op}
\end{align}
where
\begin{align}
\mathcal{P}_o &= \frac{g^2 Q^2 \omega_0^4 }{2\pi v^4 \hbar}F(\mu,T_e)  (N_e-N_L) 
\label{eq:Po} \\
F(\mu,T_e) &\equiv \int_{-\infty}^{\infty} dx |x(x-1)| [f(\hbar\omega_0 x \hspace{-2pt}- \hspace{-2pt}\hbar\omega_0) \hspace{-2pt}- \hspace{-2pt} f(\hbar\omega_0 x)].
\label{eq:FTemu}
\end{align}
$N_e = N_e(\hbar\omega_0)$ and $N_L = N_L(\hbar\omega_0)$ (see appendix \ref{sec:op_ph_nlsm}). As seen from the factor in parenthesis in Eq.(\ref{eq:Po}), $\mathcal{P}_0$ is exponentially suppressed at temperatures $k_B T_e \ll \hbar \omega_{0}$, and hence, in this regime, we expect acoustic phonon scattering to dominate. The energy is a function of $\mu$ (assumed positive)  and $T_e$, which, in turn, are functionally related due to the constant density condition. In our NLSM model the energy and electron density are 

\begin{figure}[]
\subfigure{\includegraphics[width=.5\textwidth]{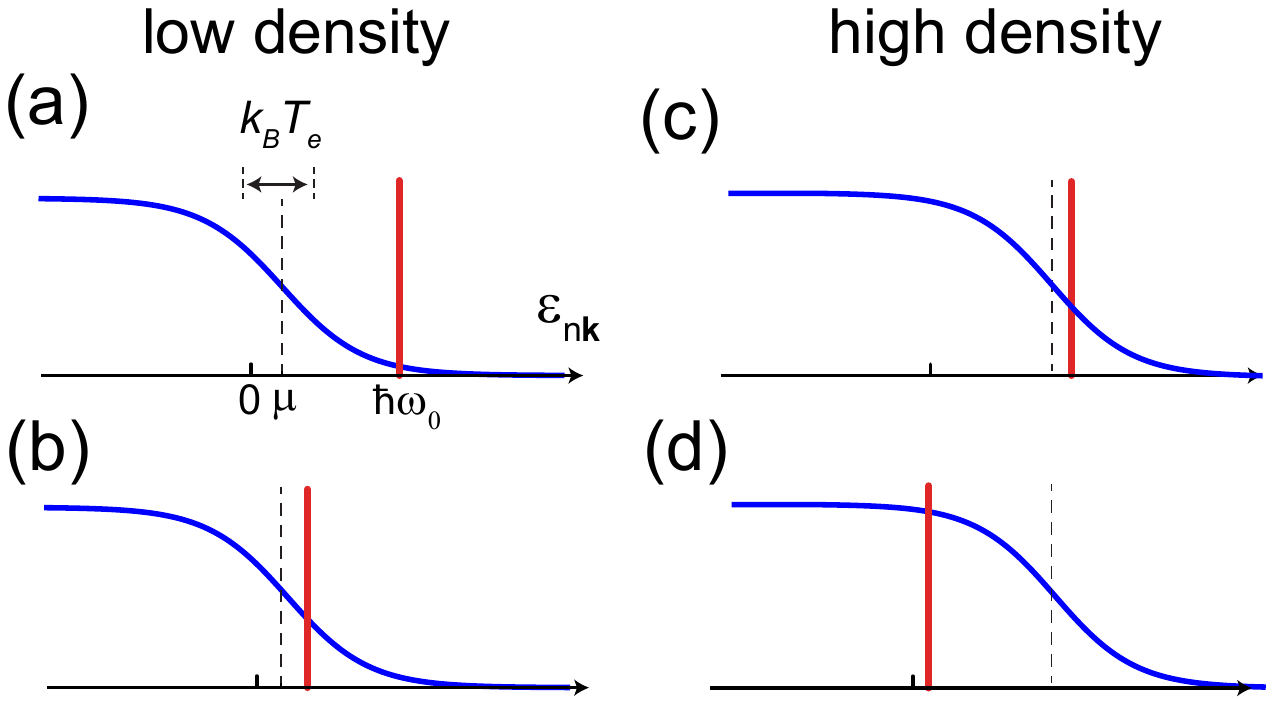}}
\caption{Instantaneous electron distribution $f(\epsilon_{n\v{k}})$ (blue curve) as a function of energy in various limits. In (a), $\mu$ (dashed line) is at the nodal line or slightly above it; that is, the system is half filled. The width of the distribution $k_B T_e$ and the optical mode energy (red line) are indicated. In (b), $\hbar\omega_0$ is of the order of $\mu$ but smaller than $k_B T_e$. (a) and (b) are \textit{low}-density regimes in the sense that $\mu\ll k_B T_e$. In (c), $\mu$ is large and of the order of $\hbar\omega_0$. Both $\mu$ and $k_B T_e$ are larger than $k_B T_e$. (d) is similar to (c) but $\hbar\omega_0\sim 0$ . (c) and (d) are called \textit{high}-density regimes in the sense that $\mu\gg k_B T_e$. $k_B T_L$ is assumed zero or close to zero.}
\label{fig:distribution_optical_ph}
\end{figure}

\begin{align}
\mathcal{E} &=\frac{1}{V}\sum_{n\v{k}} \epsilon_{n\v{k}} f_{n\v{k}}  = \frac{v\hbar Q}{2\pi} I_{2+} 
\label{eq:Enlsm} \\
n &=\frac{1}{V}\sum_{n\v{k}} f_{n\v{k}}  = \frac{ Q}{2\pi} I_{1-},
\label{eq:n}
\end{align}  
where
\begin{align}
I_{n\pm} \equiv \int_{0}^{\infty} \tilde{k}^n d\tilde{k} ~[ f_{c\tilde{k}} \pm (1-f_{v\tilde{k}} ) ],
\label{eq:Ipm_def}
\end{align}
$f_{n\tilde{k}} = [e^{\beta(nv\hbar \tilde{k} - \mu)}+1]^{-1}$, and $\beta\equiv 1/k_B T_e$.  The density is measured with respect to the nodal line.

\subsection{Low-density limit}
By low density we mean $\mu\ll k_B T_e$ and $\mu \gtrsim 0$, i.e., the system is half filled. $k_B T_L$ is assumed zero or close to zero. However, the relative size of $k_B T_e$ and $\hbar\omega_0$ gives various relaxation behaviors which we now consider in detail.

\textit{\textbf{case 1:} $\hbar \omega_{0} \gg k_B T_e\gg k_B T_L,\mu$}. Figure~\ref{fig:distribution_optical_ph}(a) illustrates this situation. The dashed line indicates the position of $\mu$ and the red line the position of $\hbar\omega_0$. $0$ indicates the position of the nodal line. In this scenario, high-energy electrons at the tail of the distribution lower their energy by emitting optical phonons and dropping to unoccupied states. A rapid thermalization among electrons (not included here) creates a new Fermi distribution at lower temperature. If the energy of the optical phonon is too large, the cooling becomes inefficient. Interestingly, the temperature at which optical-phonon cooling stops is not of the order $\hbar\omega_0$ but could be an order of magnitude smaller (see Sec.~\ref{sec:crossoverT}). 

\begin{figure}[]
\subfigure{\includegraphics[width=.48\textwidth]{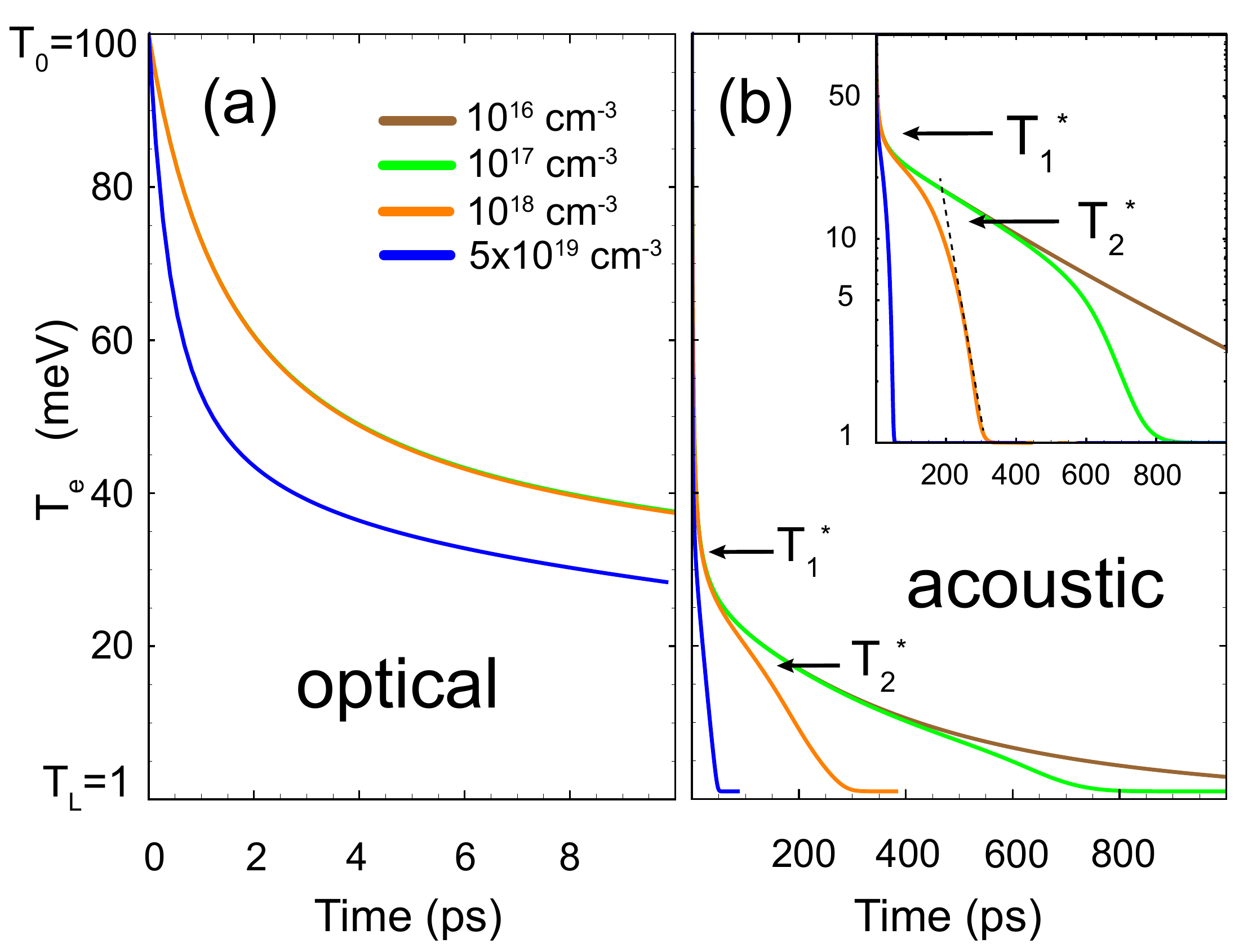}}
\caption{Numerical solution for the electron temperature relaxation in a NLSM with acoustic and optical phonon modes. (a) and (b) differ only by the time resolution of the horizontal axis. Various  electronic densities are considered. The initial temperature is $T_0=100$ meV, and the lattice temperature is $T_L=1$ meV. The optical phonon mode is at $\hbar\omega_0=300$ meV which corresponds to regime $\hbar\omega_0 \gg k_B T_e$. The rest of parameters are in Table~\ref{table:parameters_figs}. From (a) we see a time scale of $\tau_o=1-4$ ps for optical phonon relaxation, and from (b) $\tau_a=50-400$ ps, for acoustic phonon relaxation. The brown curve exhibits inverse log $\to$ exponential behaviors as $T_e$ decreases. The orange curve exhibits inverse log $\to$ linear $\to$ exponential behaviors as $T_e$ decreases. $T^{*}_1$ is the crossover temperature separating the optical- and acoustic-phonon regimes. $T^{*}_2$ marks the transition from linear to exponential.}
\label{fig:Te_relax_acop}
\end{figure}

To begin, we note that if $\mu=0$, Eq.(\ref{eq:FTemu}) is a function of $\bar{\beta}\equiv \hbar\omega_0/k_B T_e$ only and can be integrated analytically in terms of polylogarithm functions. The first two terms for large $\bar{\beta}$ are  
\begin{align}
F(0,T_e) \sim \frac{1}{6} + \frac{6 \zeta(3) }{ \bar{\beta}^3} + \cdots.~~~~~~~~ \bar{\beta} \gg 1
\label{eq:Fu_exp}
\end{align}
The total time derivative of the energy (\ref{eq:Etot}) has two contributions: one from the density $\partial \mathcal{E}/\partial \mu = 2n$ and second from the heat capacity $\partial \mathcal{E}/\partial T_e$. Now, the energy and chemical potential in the limit of low density $\mu\ll k_B T_e$ are
\begin{align}
\mathcal{E} &= \frac{Q k_B^3 T_e^3 3 \zeta{(3)}}{2\pi v^2 \hbar^2} \hspace{-2pt}+ \hspace{-2pt} \frac{Q k_B T_e \log{(2)}}{\pi v^2 \hbar^2} \mu^2 \hspace{-2pt}+ \hspace{-2pt} \cdots \mu \ll k_B T_e 
\label{eq:E_and_n1} \\
n &= \frac{Q k_B T_e \log{(2)}}{\pi v^2 \hbar^2} \mu + \cdots. ~~~~~~~~~~~\mu \ll k_B T_e
\label{eq:E_and_n2}
\end{align}  
From Eq.(\ref{eq:E_and_n2}), we see that the contribution from density vanishes in the limit $\mu \to 0$ and, hence, Eq.(\ref{eq:P_op}) becomes $(\partial \mathcal{E}/\partial T_e)(\partial T_e/\partial t) = -\mathcal{P}_o$. In the variable $\bar{\beta}$, it takes the form
\begin{align}
\frac{d \bar{\beta}}{dt} \sim \gamma_{on1} \left(\bar{\beta}^4 + 36 \zeta(3) \bar{\beta} + \cdots\right)e^{-\bar{\beta}},
\label{eq:beta_eq_neutral}
\end{align} 
where we assume $T_e \gg T_L$. The initial condition is $\bar{\beta}_0 =\hbar\omega_0/k_B T_0 \gg 1$ and  $\gamma_{on1}= g^2 Q \omega_0/v^2 \hbar^2 54 \zeta(3)$. The right hand side of Eq.(\ref{eq:beta_eq_neutral}) has a local maximum at $\bar{\beta}\sim 1$ and an exponential tail for large $\bar{\beta}$. In the limit $\bar{\beta}\gg 1$, Eq.(\ref{eq:beta_eq_neutral}) gives
\begin{align}
\bar{\beta} \sim \log{(\gamma_{on1}\bar{\beta}_{0}^{4}t + e^{\bar{\beta}_0})} + 4\log{\frac{T_0}{T_e}}, 
\label{eq:op_ph_Te_relax_neutral0}
\end{align}
or the inverse log 
\begin{align}
k_{B} T_e \sim \frac{\hbar \omega_{0}}{\log(\gamma_{on1}\bar{\beta}_{0}^{4} t + e^{\bar{\beta}_0})}.
\label{eq:op_ph_Te_relax_neutral}
\end{align}
In the last expression, we drop a small (for $T_e \lesssim T_0$) logarithmic correction. The relaxation time scale is $e^{\bar{\beta}_0}/\gamma_{on1}\bar{\beta}_{0}^{4}$. Note that $\bar{\beta}$ increases monotonically with time (as $T_e$ decreases) and, hence, it is enough to request $\bar{\beta}_0\gg 1$. Figure~\ref{fig:Te_relax_acop}(a)  shows a numerical (see Sec.~\ref{sec:numerical_sol}) example of optical phonon relaxation in the low-density regime (brown curve) with $\bar{\beta}_0 =3\gg 1$ . The temperature relaxation is approximately given by Eq.(\ref{eq:op_ph_Te_relax_neutral}).

\begin{figure}[]
\subfigure{\includegraphics[width=.48\textwidth]{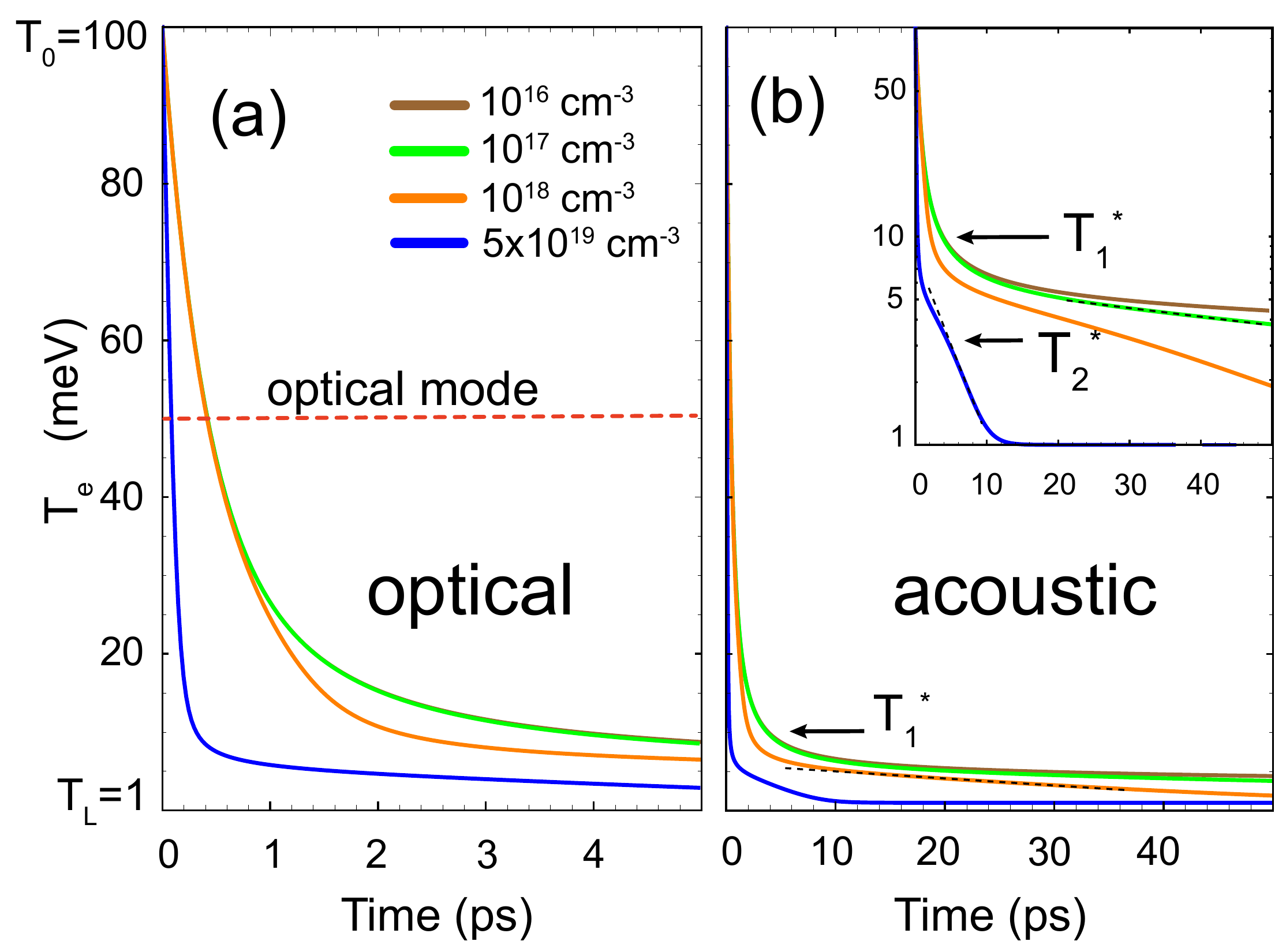}}
\caption{Same as in Fig.~\ref{fig:Te_relax_acop}, but with $\hbar\omega_{0}=50$ meV, i.e., $\hbar\omega_0 < k_B T_0$. The time scales are $\tau_a=10-50$ ps and $\tau_o=.5-1$ ps for acoustic and optical phonon relaxations, respectively. The low density green curve exhibits transition from exponential $\to$ exponential as $T_e$ decreases. The high density blue curve exhibits linear$\to$linear$\to$exponential behavior as $T_e$ decreases.}
\label{fig:Te_relax_ac_op_omega0_small}
\end{figure}

\textit{\textbf{case 2:} $k_B T_e\gg \hbar \omega_{0}, k_B T_L,\mu$}. Figure~\ref{fig:distribution_optical_ph}(b) illustrates this situation. In this regime, many electrons have energy above the optical phonon mode and, hence, we expect enhanced optical phonon emission and subsequent fast relaxation. Indeed, we find an exponential relaxation in this regime. To see this, note that for $k_B T_e \gg \hbar \omega_{0}$ (or $\bar{\beta} \ll 1$), Eq.(\ref{eq:FTemu}) is approximated by  
\begin{align}
F(0,T_e) = \frac{\pi^2}{3\bar{\beta}^2} -\frac{1}{6} + \frac{\bar{\beta}}{12} - \frac{\bar{\beta}^3}{360}+ \cdots. ~~~\bar{\beta} \ll 1
\label{eq:Fon2}
\end{align}
If, in addition, $T_e \gg T_L$, then $N_e(\hbar\omega_0)- N_L(\hbar\omega_0) \sim k_B T_e/\hbar\omega_0 - k_B T_L/\hbar\omega_0 \sim 1/\bar{\beta}$ and to leading order the relaxation is exponential 
\begin{align}
\frac{d T_e}{d t} = -\gamma_{on2} T_e.
\label{eq:Pon2} 
\end{align}
Here, $\gamma_{on2}= g^2 Q \omega_0 \pi^2 / v^2 \hbar^2 27 \zeta (3)$ and the relaxation time scale is $1/\gamma_{on2}$. Figure~\ref{fig:Te_relax_ac_op_omega0_small}(a) shows the relaxation of a NLSM with optical phonon mode energy half of the initial temperature of the electron ensemble, $\bar{\beta}_0 =1/2$ (brown/green curves). The temperature relaxation is approximately exponential in agreement with Eq.(\ref{eq:Pon2}).  

\subsection{High-density limit}
\label{sec:nlsm_op_doped}
By high density we mean $\mu\gg k_B T_e$. Since $\mu$ is bounded from above by the Fermi energy $\epsilon_F=\mu(T_e=0)$, high density means $\epsilon_F \gg k_B T_e$. However, the relative size of $k_B T_e$ and $\hbar\omega_0$ gives various relaxation behaviors which we now consider in detail.

\textit{\textbf{case 1:} $\mu, \hbar\omega_0 \gg k_B T_e \gg  k_B T_L$}. Figure~\ref{fig:distribution_optical_ph}(c) illustrates this situation. To start, note that for $\mu \gg k_B T_e$, Eq.(\ref{eq:FTemu}) becomes
\begin{align}
F(\mu,T_e) \sim \left(\frac{1}{6} + \frac{2 \bar{\mu}^3}{3}\right) + \frac{2\pi^2 \bar{\mu}}{3 \bar{\beta}^2} + \cdots,~~\mu \gg k_B T_e
\label{eq:Fu_exp2}
\end{align}
which is obtained from the Sommerfeld expansion in variables $\bar{\mu}\equiv \mu/\hbar\omega_0$ and $\bar{\beta}\equiv \hbar\omega_{0}/k_B T_e$, and is valid for $\bar{\mu} \bar{\beta}  \gg 1$ (or equivalently, $\mu\gg k_B T_e$). In this limit, the energy and electron density become 
\begin{align}
\mathcal{E} &= \frac{Q}{6\pi v^2 \hbar^2} \big(\mu^3 + \frac{\pi^2 \mu}{\beta^2} +\cdots \big), ~~~~~\mu \gg k_B T_e 
\label{eq:E_largemu} \\
 n &= \frac{Q}{4\pi v^2 \hbar^2} \big(\mu^2 + \frac{\pi^2}{3 \beta^2} +\cdots \big), ~~~~~\mu \gg k_B T_e
\label{eq:n_largemu}
\end{align}
and hence $\mu (T_e)= \epsilon_F - \pi^2 k_B^2 T_e^2/6\epsilon_F + \cdots$. Also, if  $\bar{\beta} \gg 1$, then $N_e - N_L \sim e^{-\hbar\omega_0/k_B T_e} - e^{-\hbar\omega_0/k_B T_L} \sim e^{-\bar{\beta}}$ and Eq.(\ref{eq:P_op}) becomes

\begin{align}
\frac{\partial \bar{\beta}}{\partial t} = \tilde{\gamma}_{od1}\bar{\beta}^3 (\frac{1}{6} + \frac{2 \bar{\mu}^3}{3} + \frac{2\pi^2 \bar{\mu}}{3\bar{\beta}^2}+ \cdots)e^{-\bar{\beta}}.
\end{align}
Now, if $\bar{\mu} \bar{\beta} \gg \pi$, the leading behavior is an inverse log function 
\begin{align}
k_B T_e = \frac{\hbar \omega_0}{\log(\gamma_{od1}\bar{\beta}^{3}_{0} t + e^{\bar{\beta}_0})},
\label{eq:nlsm_op_doped1}
\end{align}
where $\gamma_{od1}= 3 g^2 Q \omega_0^2 (1/6 + 2\bar{\mu}^{3}/3) /\pi^2 v^2 \mu \hbar$. The relaxation time scale $e^{\bar{\beta}_0}/\gamma_{od1}\bar{\beta}^{3}_{0}$ goes as $\sim n^{1/2}$ for $\bar{\mu}\ll 1$ and as $\sim n^{-1}$ for $\bar{\mu} \gg 1$; $n$ is the electron density. Note that the time scale increase or decrease with increasing electron density. Figure\ref{fig:Te_relax_acop}(a) illustrates the high-density regime with parameters $\bar{\beta}_0 =3$, $\bar{\mu} \sim \epsilon_F/\hbar\omega_0 \sim 1$ (blue curve) . The relaxation is approximately given by Eq.(\ref{eq:nlsm_op_doped1}).

\textit{\textbf{case 2:} $\mu \gg k_B T_e \gg \hbar\omega_0, k_B T_L$}. Figure \ref{fig:distribution_optical_ph}(d) illustrates this situation. Since $\mu \gg k_B T_e$ we can use Eq.(\ref{eq:Fu_exp2}). In addition, if $\bar{\beta} \ll 1$ and $T_e \gg T_L$, then $N_e-N_L \sim 1/\bar{\beta}$ and Eq.(\ref{eq:P_op}) becomes
\begin{align}
\frac{d\bar{\beta}}{dt} = \tilde{\gamma}_{od2}\bar{\beta}^{2}  (\frac{1}{6} + \frac{2 \bar{\mu}^3}{3} +  \frac{2 \pi^2 \bar{\mu}}{3\bar{\beta}^2} \cdots ).
\end{align}
Now, if $ \bar{\mu}\bar{\beta} \gg \pi$, the leading behavior is linear 
\begin{align}
k_B T_e =k_B T_0 - \hbar\omega_0 \gamma_{od2} t,
\label{eq:relax_nlsm_over_case2}
\end{align}
where $\gamma_{od2}=3 g^2 Q \omega_0^2 (1/6 + 2\bar{\mu}^3/3)/ \pi^2 v^2 \hbar \mu$ goes as $\sim n^{-1/2}$ for $\bar{\mu}\ll 1$ or as $\sim n$ for $\bar{\mu}\gg 1$. The rate can increase or decrease with increasing electron density. Figure \ref{fig:Te_relax_ac_op_omega0_small}(a) illustrates the high-density regime with parameters $\bar{\beta}_0 =1/2$ (blue or orange curve). The relaxation is approximately linear for temperatures above $\sim 40$ meV.

\begin{figure}[]
\subfigure{\includegraphics[width=.44\textwidth]{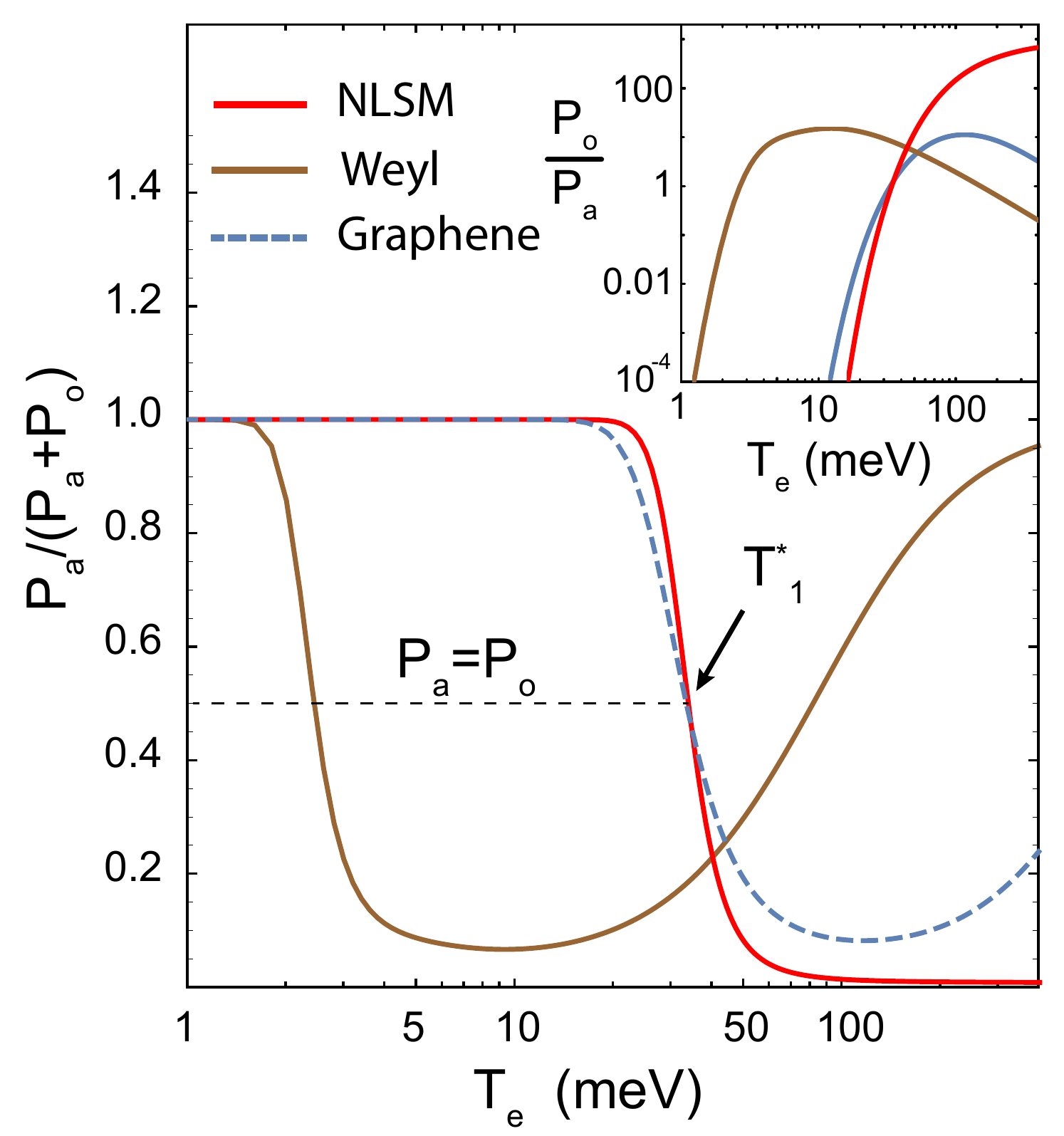}}
\caption{Percentage of acoustic phonon cooling power as a function of electron temperatures for NLSM, graphene and a Weyl SM. Parameters are given in Table~\ref{table:parameters_figs}}
\label{fig:ZrSiS-TaAs-graphene_all}
\end{figure}

\section{Acoustic phonon relaxation}
\label{acustic_ph_cooling_power}
We now calculate the temperature relaxation due to acoustic phonons in the low and high density regimes.  To obtain analytical expressions, we evaluate Eq.(\ref{eq:P_gen}) to lowest order in $c/v \ll 1$ where $c$ is the sound speed, assumed isotropic. In this limit, acoustic phonon scattering is quasielastic and we obtain (see appendix \ref{sec:acoustic_ph_nlsm})
\begin{align}
\frac{d\mathcal{E}}{dt} &= - \mathcal{P}_a,
\label{eq:P_ac}
\end{align}
where
\begin{align}
\mathcal{P}_a &= \frac{D^2 Q^4 k_{B}}{4\pi \rho v^2 \hbar}I_{1+}(T_e- T_L). 
\label{eq:Pa}
\end{align}

\subsection{Low-density limit}
Setting $\mu=0$ and using Eqs.(\ref{eq:E_and_n1}), (\ref{eq:E_and_n2}), and (\ref{eq:Pa}) we obtain
\begin{align}
\frac{d T_e}{dt} = -\gamma_{an} (T_e - T_L),
\label{eq:ph_relax_neutral}
\end{align}
where $\gamma_{an}=\pi^2 D^2 Q^3/\rho v^2 \hbar 108\zeta(3)$. Hence, the temperature relaxes exponentially with time scale $1/\gamma_{an}$. This should be compared with the slower $\sim 1/\sqrt{t}$ power law found in graphene \cite{Bistritzer2009}, which has nodal points instead of nodal lines and the even slower $\sim 1/\sqrt[3]{t}$ power law found in Weyl nodes \cite{Lundgren2015}. Figure \ref{fig:Te_relax_acop}(b) shows an example of the low-density relaxation in of a NLSM (brown curve). The exponential dependence is more evident in the log plot inset.

\subsection{High-density limit}
\label{sec:doped}
In this limit, we can use Eqs.(\ref{eq:E_largemu}), (\ref{eq:n_largemu}), and (\ref{eq:Pa}) to obtain 
\begin{align}
\frac{d}{dt} (k_B T_e) = - \gamma_{ad} \frac{1}{T_e} (T_e - T_L), 
\label{eq:doped_rel}
\end{align}
where $\gamma_{ad} = 3 D^2 Q^3\epsilon_F /4 \pi^2 \rho \hbar v^2$. We find linear relaxation for $T_e \gg T_L$ with a relaxation \textit{rate} that scales as $\gamma_{ad} \sim n^{1/2}$ with electron density. This is different from graphene where the relaxation rate goes as  $ \sim n^{3/2}$ \cite{Bistritzer2009}. Finally, the relaxation is exponential for $T_e \gtrsim T_L$ with time scale $T_L/\gamma_{ad}$.

Figure \ref{fig:Te_relax_acop}(b) shows the temperature relaxation in the high-density regime of a NLSM (orange or blue curve). The appearance of a linear relaxation in the temperature range $T^{*}_1 >T > T^{*}_2$ is evident in the figure and is in agreement with Eq.(\ref{eq:doped_rel}).  $T^{*}_1$ is the crossover temperature between optical relaxation to acoustic relaxation (see Sec.~\ref{sec:crossoverT}). $T^{*}_2$ marks the regime transition from linear to exponential relaxation.  There is an exponential relaxation for $T < T^{*}_2$.

\begin{table}
\caption{Parameters of a prototypical NLSM from Ref.\cite{Rudenko2020}, a Weyl SM, and doped graphene. The electron-phonon coupling constants for NLSMs are unknown. For a Weyl SM, we take the known parameters of TaAs \cite{Armitage2018}}
\label{table:parameters_figs}
\begin{center}
\begin{tabular}{l c c c} 
 \hline\hline
    																			& ~~~NLSM~~~        &  ~~~Weyl~~~     &  ~~~Graphene~~~        \\ [0.5ex] 
 \hline 
 $\rho$ ~~(kg/m$^3$,kg/m$^2$ )    					    & 4.86       &  12.03               &  3.7$\times$ 10$^{-7}$      \\ 
 $n$ ~~(cm$^{-3}$,cm$^{-2}$)     						    & 5$\times$10$^{19}$ &   10$^{14}$   & 10$^{13}$   \\  
 $\epsilon_F$ ~~(meV)     						          & 300          &  6                  & 370    \\  
 $Q$ ~~(1/\AA)           		  			          	& 0.3           &  -                  & -           \\  
 $\hbar\omega_0$ (meV)           		  			  	& 300          &  31                 & 196          \\  
 $D$ ~~(eV)                      						  	& 5            &  4                  & 20           \\ 
 $v$  ~~(10$^6$ m/s)               					  	& 1            &  .5                 & 1          \\ 
 $a$  ~~(\AA)                   						    & 4            &  6.3                & 1.42         \\ 
 $g$ ~~($v \hbar^2/a^2\sqrt{\rho\hbar\omega_0}$)& 56.6         &  4.2                & 1.4           \\  
 $T_L$ ~(meV)                                 & 1             & 1                     &  1         \\
 \hline\hline
 \end{tabular}
\end{center}
\end{table}

\section{Crossover temperature}
\label{sec:crossoverT}
The temperature $T^{*}_1$ at which 
\begin{align}
\mathcal{P}_a(T_1^{*}) = \mathcal{P}_o(T_1^{*}),
\end{align}
defines the boundary between the regime dominated by optical phonons ($>T_{1}^{*}$) and the regime dominated by acoustic phonons  ($<T_{1}^{*}$). If the initial temperature is $T_0>T_{1}^{*}$, the decay is initially fast followed by a slower decay. If $T_0 < T_{1}^{*}$, the decay is slower and the electron plasma is longer lived. Hence, an estimate of the crossover temperature is important for experimental design, device design, or as a control knob in probing  microscopic mechanisms in materials. We consider two limiting cases in detail.

\subsection{Low-density limit}

\textit{\textbf{case 1:} $\hbar\omega_0 \gg k_B T_e \gg k_B T_L,\mu$}. From Eqs.(\ref{eq:Po}), (\ref{eq:Fu_exp}), and (\ref{eq:Pa})  we find that $\bar{\beta}^{*} =\hbar\omega_0/k_B T_1^{*}$ satisfies
\begin{align}
\alpha_1 \equiv \frac{\pi^2 D^2 Q^2 \hbar}{2\rho g^2 \omega_0} = \big[(\bar{\beta}^{*})^{3} + 36\zeta(3)\big] e^{-\bar{\beta}^{*}}.
\label{eq:paequalpo_neutral1}
\end{align} 
The left hand side is a constant that depends on materials' properties. The right hand side depends on $\bar{\beta}^{*}$. The intersection of the two gives the solution. Note that if $\alpha_1 <  36\zeta(3)$ and $\bar{\beta}^{*} \gg 1$, there is a solution.

\textit{\textbf{case 2:} $k_B T_e \gg  \hbar\omega_0,  k_B T_L,\mu$}. From Eqs.(\ref{eq:Po}), (\ref{eq:Fon2}), and (\ref{eq:Pa}), we find that $\bar{\beta}^{*} =\hbar\omega_0/k_B T_1^{*}$ satisfies 

\begin{align}
\alpha_1 = 2\pi^2 - \bar{\beta}^{*2} + \frac{\bar{\beta}^{*3}}{2}  + \cdots.
\label{eq:paequalpo_neutral2}
\end{align} 
There is a solution when $\alpha_1 < 2\pi^2$ and $\bar{\beta}^{*} \ll 1$.

\subsection{High-density limit}

\textit{\textbf{case 1:} $\mu,\hbar\omega_0 \gg k_B T_e \gg k_B T_L$}. From Eq.(\ref{eq:Po}), (\ref{eq:Fu_exp2}), and (\ref{eq:Pa}), we find that $\bar{\beta}^{*} =\hbar\omega_0/k_B T_1^{*}$ satisfies 
\begin{align}
\alpha_2 \equiv \frac{3 D^2 Q^2 \mu^2}{2\rho g^2 \omega_0^3 \hbar} = \big[\bar{\beta}^{*}(1+4\bar{\mu}^3) + 4\pi^2 \frac{\bar{\mu}}{\bar{\beta}^{*}} \big] e^{-\bar{\beta}^{*}}.
\label{eq:paequalpo_doped}
\end{align} 
Using the parameters in Table~\ref{table:parameters_figs}, $\alpha_2 = 0.0042$ and the intersection with the term on the right is at $k_B T^{*}_1 = \hbar\omega_0/\bar{\beta}^{*}= 34$ meV, which is close to the full numerical solution $k_B T^{*}_1 =  35$ meV shown in Fig.~\ref{fig:ZrSiS-TaAs-graphene_all}.

\textit{\textbf{case 2:} $\mu \gg k_B T_e \gg \hbar\omega_0, k_B T_L$}. From Eqs.~\ref{eq:Po}, \ref{eq:Fu_exp2}, \ref{eq:Pa} and $N_e - N_L\sim 1/\bar{\beta}$, we find 

\begin{align}
 k_B T_1^{*} = \frac{D Q \hbar}{2\pi g }\bigg(\frac{3\mu}{2\rho}\bigg)^{1/2}\sim n^{1/4}.
\end{align}

\section{Acoustic- and optical-phonon relaxation: numerical solution}
\label{sec:numerical_sol}
In this section, we solve numerically for $T_e(t)$ and $\mu(t)$ in the presence of acoustic and optical phonons. The basic equations are  
\begin{align}
\frac{d\mathcal{E}}{dt} &=  - \mathcal{P}_a - \mathcal{P}_o, 
\label{eq:Pac_op}\\
n &=\frac{1}{V}\sum_{n\v{k}} f_{n\v{k}}.
\end{align}
The first equation gives the energy dynamics and the second gives the constant electron density condition. $\mathcal{P}_a$ and $\mathcal{P}_o $ are the acoustic- and optical-phonon cooling power in Eqs.(\ref{eq:Po}) and (\ref{eq:Pa}), and $\mathcal{E}$ is the energy
\begin{align}
\mathcal{E} &=\frac{1}{V}\sum_{n\v{k}} \epsilon_{n\v{k}} f_{n\v{k}}.
\label{eq:Enlsm_num_sec} 
\end{align}  
More explicitly, Eq.(\ref{eq:Pac_op}) becomes 
\begin{align}
-3Q &T_e v \hbar (\frac{\partial}{\partial t} \frac{1}{T_e}) I_{2+} + 2Q T_e (\frac{\partial}{\partial t}\frac{\mu}{T_e}) I_{1-} = \nn \\
&-\frac{D^2 Q^4 k_{B}}{4\pi \rho v^2 \hbar}I_{1+}(T_e- T_L) -\frac{g^2 Q^2 \omega_0^4}{2\pi v^4 \hbar}  F~  (N_e-N_L),
\end{align}
and the constant density condition
\begin{align}
n & = \frac{ Q}{2\pi} I_{1-}.
\end{align}

\begin{table}
\caption{Cooling power and heat capacity scaling with temperature for Dirac materials (see Sec.~\ref{sec:conclusions}). $\mathcal{D}$ spatial dimension, $d$ nodal manifold dimension,  $dim~\mathcal{C}$ scale dimension of the heat capacity, and $dim~\mathcal{P}_a$ scale dimension of acoustic phonon cooling power.}
\begin{center}
\begin{tabular}{ l c c c } 
 \hline\hline
    															   	& ~~~~~NLSM~~~~~        &  ~~~~~Weyl~~~~~          &  ~~~~~Graphene~~~~~   \\ [0.5ex] 
 \hline 
 $\mathcal{D}$          					    & 3           &  3             &  2           \\ 
 $d$                  						    & 1           &  0             & 0            \\  
 $2[q]$          						          & 0           &  2             & 2            \\  
 $dim~ \mathcal{C}$      		        	& 2           &  3             & 2            \\  
 $dim~ \mathcal{P}_a-1$        			  & 2           &  6             & 4            \\  
 $\kappa$                             &\hspace{2pt} 0$^a$  & \hspace{2pt} 3$^b$   & \hspace{3pt} 2$^c$    \\ 
 \hline\hline
 \end{tabular}
\label{table:dim_nlsm}
\end{center}
\begin{FlushLeft}
$^a$Equation (\ref{eq:ph_relax_neutral}). \\
$^b$Equation (\ref{eq:weyl_acoustic_neutral}). \\
$^c$Equation (\ref{eq:graphene_acoustic_neutral}).
\end{FlushLeft}
\end{table}

The initial conditions are $\mu(t=0)=\mu_0$, which is given by the electron density and $T_{e}(t=0)=T_0$.  Figure \ref{fig:Te_relax_acop} shows the electron temperature as a function of time for a NLSM with parameters given in Table~\ref{table:parameters_figs}. The initial temperature is $T_0=100$ meV $= 1200$ K. Note a sharp change in behavior at temperature $T^{*}_1$. For $T_e > T^{*}_1 \sim 35$ meV, there is a fast relaxation associated with optical phonons. At these temperatures, $\mathcal{P}_a \ll \mathcal{P}_o$  (see Fig.~\ref{fig:ZrSiS-TaAs-graphene_all}). For $T_e < T^{*}_1$, acoustic phonons saturate the cooling power and dominate the relaxation process. In the optical-phonon regime, the relaxation time scale is picoseconds, but in the acoustic-phonon regime, the relaxation time scale is nanoseconds. As expected, the optical phonon relaxation is faster than acoustic phonon relaxation \cite{Gantmakher1987}.

Note that the higher the density, the faster the relaxation is. In the optical-phonon regime at any density, the relaxation is of the inverse log form, as indicated in Eq.(\ref{eq:op_ph_Te_relax_neutral}) or Eq.(\ref{eq:nlsm_op_doped1}), even though $\bar{\beta}_0$ is only 3. In the acoustic-phonon, low-density regime, the relaxation is exponential for $T_e < T^{*}_1$ [brown curve Eq.(\ref{eq:ph_relax_neutral})] and linear-exponential (orange curve) in the high-density regime [Eq.~\ref{eq:doped_rel}]. The relaxation time scale depends weakly on density in the optical-phonon regime and strongly in the acoustic-phonon regime. Although not shown, the chemical potential increases monotonically towards $\epsilon_F$ as time increases. 

In the optical-phonon regime, $\mathcal{P}_a$ is expected to be low. What is interesting is that $\mathcal{P}_a$ in NLSMs is lower than in a Weyl SMs and graphene (see Fig.~\ref{fig:ZrSiS-TaAs-graphene_all}). In fact, in NLSMs, there is a clean separation of scattering processes at $T_1^{*}$, which justifies considering optical and phonon scattering separately in Secs.~\ref{sec:optica_ph} and \ref{acustic_ph_cooling_power}. 

Figure\ref{fig:Te_relax_ac_op_omega0_small} shows the relaxation of a NLSM in the regime $k_B T_e\gg \hbar\omega_{0}$.  Since emission of optical phonons is readily available, we expect a faster relaxation, as shown.

\section{Acoustic-phonon relaxation in Dirac materials}
\label{sec:scaling_arg}
The role of the nodal line is most critical in the acoustic-phonon, low density regime. This is expected because at low density ($\mu =0$) acoustic phonons probe quasiparticles near the nodal line in NLSMs or nodal points in Weyl SMs or graphene. In this regime, the NLSM has exponential relaxation, Weyl SMs have $\sim 1/\sqrt[3]{t}$ relaxation, and graphene has $\sim 1/\sqrt{t}$. To understand this, write we the energy relaxation equation as $\mathcal{C} d T_e/dt =- \mathcal{P}_a$, where  the heat capacity $\mathcal{C} = d \mathcal{E}(\mu(T_e),T_e)/d T_e $. By power counting, we can write
\begin{align}
\frac{d T_e}{dt} \sim - T_e^{\kappa} (T_e - T_L) 
\end{align} 
with $\kappa = dim~ (\mathcal{P}_a -1) - dim ~\mathcal{C}$ or
\begin{align}
\kappa = (2(\mathcal{D}-d) - 1 + 2[q]- 1) - (\mathcal{D}-d +a -1).
\label{eq:kappa}
\end{align} 
To obtain $dim~ \mathcal{C}$, we can inspect Eq.(\ref{eq:Enlsm}). First subtract $d$ nodal (frozen) directions from $\mathcal{D}$ spatial dimensions and add the power of the quasiparticle dispersion $k^{a}$ (1 in this case). The temperature derivative of the energy gives the last -1. The electronic heat capacity scales as $T_e$ with a power of 3 in Weyl nodes and with a power of 2 in graphene and NLSMs. In this sense, NLSMs and graphene are similar. 

To obtain $dim~ \mathcal{P}_a-1$, we can inspect Eq.(\ref{eq:P_gen}). We have $2(\mathcal{D}-d)$ powers from the integrals, subtract 1 from the energy conservation, and add 2 if momentum transfer $[q]$ is not constrained, i.e., in Weyl nodes and graphene. Finally, subtract 1 from the Fermi function differences. The last two steps occur because we expanded the integrals to lowest order in $c/v$. The role of the nodal line is now explicit. From Eq.(\ref{eq:kappa}), $dim~ \mathcal{P}_a-1 =6$ for Weyl SMs, $dim~ \mathcal{P}_a-1 =4$ for graphene and $dim~ \mathcal{P}_a-1 =2$ for NLSMs  (see Table~\ref{table:dim_nlsm}). It is lowest in NLSMs. There are two consequences: (a) a large decrease in $\mathcal{P}_a$ in a NLSM at high temperature compared with Weyl SMs and graphene (see Fig.~\ref{fig:ZrSiS-TaAs-graphene_all}), and (b) at low temperatures, it gives an exponential temperature relaxation in NLSMs and a slower power law in graphene and Weyl SMs.

\section{Discussion and conclusions}
\label{sec:conclusions}
We calculated the electron temperature relaxation as a function of time after a sudden excitation. We use a two-band model of a NLSM with a single acoustic branch and a single optical branch. Among our findings, we point out that acoustic-phonon relaxation is exponential, which is similar to standard 3D metals~\cite{Allen1987} but different from Weyl SM and graphene. This is because a particular combination of FS volume and topology in NLSMs. Other relaxation behaviors were obtained depending on initial conditions and density and were summarized in Table~\ref{table:summary}. Our results point to possible ways to engineer thermalization timescales e.g., by tuning either density, spatial dimension or the shape of the FS.  

We assumed that phonons of large momenta $\sim 2Q$ can be thermally excited, and this puts a lower bound on $T_e$, $k_B T_e > \hbar c 2Q \equiv k_B T_{BG}$. For for ZrSiS~\cite{Rudenko2020} with $c= 4.3-6.8$ km/s and $Q \sim .3$ \AA$^{-1}$ we obtain $k_B T_{BG} = 17-26$ meV = $197-300$ K. For $PbTaSe_2$~\cite{Chang2016,Bian2016} with roughly the same sound speed and half $Q$ we obtain $k_B T_{BG} = 98-150$ K.

We have assumed particle-hole symmetry, time reversal symmetry, and a highly symmetric nodal line. Our conclusions will not change for small deviations of these symmetries. If there is more than one ring or if we consider the spin of the electrons, the total cooling power is multiplied by the number of equivalent rings and by the spin degeneracy, but $T_e(t)$ will not change. In this work we assumed a clean system, future work is needed to understand impurities effects.

\section{acknowledgments}
B.M.F. acknowledges support from NSF Grant DMR-2015639 and DOE-NERSC under Contract DE-AC02-05CH11231. M.N. is supported by the Air Force Office of Scientific Research under Award No. FA9550-17-1-0415 and the Air Force Office of Scientific Research MURI (FA9550-20-1-0322).

\appendix

\section{Phonon cooling power}
\label{sec:ph_coolingpower}

From Eq.(\ref{eq:Ptot}), (\ref{eq:dfdt}), and (\ref{eq:Wmatrix_el}) we obtain 

\begin{align}
\mathcal{P} &= \frac{1}{V}\sum_{n\v{k}} \sum_{m\v{p}} (\epsilon_{n\v{k}} - \epsilon_{m\v{p}}) f_{n\v{k}} (1- f_{m\v{p}}) W_{n\v{k},m\v{p}} \\
&=\mathcal{P}_{1} + \mathcal{P}_{2},
\end{align}
where
\begin{align}
\mathcal{P}_{1} = \frac{2\pi}{\hbar} \frac{1}{V}\sum_{n\v{k}}& \sum_{m\v{p}} (\epsilon_{n\v{k}} - \epsilon_{m\v{p}}) (f_{n\v{k}} - f_{m\v{p}}) M_{\v{q}} N_{L}\times \nn  \\
& ~~~~~~~~~~~~~\delta(\epsilon_{n\v{k}} - \epsilon_{m\v{p}} - \hbar \omega_{\v{q}}) 
\label{eq:pa1_pa2_0} \\
\mathcal{P}_{2} = \frac{2\pi}{\hbar}\frac{1}{V} \sum_{n\v{k}}& \sum_{m\v{p}} (\epsilon_{n\v{k}} - \epsilon_{m\v{p}}) f_{n\v{k}} (1 - f_{m\v{p}}) M_{\v{q}}\times \nn \\
& ~~~~~~~~~~~~~\delta(\epsilon_{n\v{k}} - \epsilon_{m\v{p}} - \hbar \omega_{\v{q}}).
\label{eq:pa1_pa2}
\end{align}
In obtaining Eqs.(\ref{eq:pa1_pa2_0}) and (\ref{eq:pa1_pa2}), we assumed $\omega_{-\v{q}} =\omega_{\v{q}}$, which holds for time reversal symmetric systems or inversion symmetric systems. Using the identity
\begin{align}
f_{n\v{k}}(1-f_{m\v{p}})&\delta(\epsilon_{n\v{k}}-\epsilon_{m\v{p}}-\hbar\omega_{\v{q}}) = \nn \\
- (f_{n\v{k}}-&f_{m\v{p}}) N_{e}(\hbar\omega_{\v{q}})\delta(\epsilon_{n\v{k}}-\epsilon_{m\v{p}}-\hbar\omega_{\v{q}}),
\label{eq:identity}
\end{align}
$\mathcal{P}_{2}$ becomes

\begin{align}
\mathcal{P}_{2} &= -\frac{2\pi}{\hbar} \frac{1}{V}\sum_{n\v{k}} \sum_{m\v{p}} (\epsilon_{n\v{k}} - \epsilon_{m\v{p}}) ( f_{n\v{k}} - f_{m\v{p}}) M_{\v{q}} N_{e} \nn  \\
&~~~~~~~~~~~~~~~~~~~~~~~~~~~~~~~~~ \times \delta(\epsilon_{n\v{k}} - \epsilon_{m\v{p}} - \hbar \omega_{\v{q}}) \nn \\
&\equiv-\mathcal{P}_{1} (L\to e).
\label{eq:pa2_pa1_relation}
\end{align}
This relation (noted in Ref.~\cite{Viljas2010}) means Eq.(\ref{eq:Ptot}) in the main text becomes 
\begin{align}
\mathcal{P} &= \frac{2\pi}{V\hbar} \sum_{n\v{k}m\v{p} } (\epsilon_{n\v{k}} - \epsilon_{m\v{p}}) (f_{n\v{k}}   -  f_{m\v{p}})  (N_L   -   N_e ) M_{\v{q}} \times\nn \\
&~~~~~~~~~~~~~~~~~~~~~~~\delta(\epsilon_{n\v{k}}   -  \epsilon_{m\v{p}}  -  \hbar\omega_{\v{q}}),
\label{eq:P_gen2}
\end{align}
which is Eq.(\ref{eq:P_gen}) in the main text.

\section{Optical phonon cooling in NLSMs}
\label{sec:op_ph_nlsm}
We start from Eq.(\ref{eq:pa1_pa2_0}) and set $M_{\v{q}}=g^2/V$, $\omega_{\v{q}}=\omega_{0}$ to obtain

\begin{align}
\mathcal{P}_{o1} \hspace{-2pt}=\hspace{-2pt} 2\pi g^2 N_{L} \omega_{0}\frac{1}{V^2} \sum_{n\v{k}} \sum_{m\v{p}} (f_{n\v{k}}\hspace{-2pt}-\hspace{-2pt}f_{n\v{p}}) \delta(\epsilon_{n\v{k}}\hspace{-2pt}-\hspace{-2pt}\epsilon_{n\v{p}}\hspace{-2pt}-\hspace{-2pt}\hbar\omega_{0}).
\end{align}
Now, we write the sums over momenta as integrals in cylindrical coordinates. Then define a new variable $k'\equiv k-Q$. The integral over $k'$ is now from $-Q$ to $\infty$. Next, extend the integration of $k'$ to the whole real line and parameterize the plane $k'$-$k_z$ in  polar coordinates $\tilde{k},\phi_{\tilde{k}}$ to obtain

\begin{align}
\mathcal{P}_{o1} &=  g^2 N_{L} \omega_{0} \frac{1}{(2\pi)^3} \sum_{nm } \int \tilde{k} d\tilde{k} d\phi_{\tilde{k}} \int \tilde{p} d\tilde{p} d\phi_{\tilde{p}} \nn \\
 &\times (\tilde{k}\sin\phi_{\tilde{k}} + Q) (\tilde{p}\sin\phi_{\tilde{p}} + Q) \big[f(nv\hbar \tilde{k})-f(mv\hbar \tilde{p})\big] \nn\\
&~~~~~~~~\delta(\epsilon_{n\tilde{k}}-\epsilon_{n\tilde{p}}-\hbar\omega_{0}) \\
&=  g^2 N_{L} \omega_{0} \frac{Q^2}{2\pi v\hbar} \sum_{nm } \int \tilde{k} d\tilde{k} \int \tilde{p} d\tilde{p} \nn \\
 & \times \big[f(nv\hbar \tilde{k})-f(mv\hbar \tilde{p})\big] \delta(n\tilde{k}-m\tilde{p}-\omega_{0}/v),
\end{align} 
where in the last equality we use the fact that the delta function fixes $m\tilde{p} = n\tilde{k} + \omega_{0}/v$ and that the Fermi-function difference is very small for $\tilde{k} \gg  \omega_0/v$. This justifies extending the limits of integrals to the whole real line as long as $Q \gg \omega_0/v,k_B T_e/v\hbar$. After some algebra
\begin{align}
\mathcal{P}_{o1} &= -  \frac{g^2\omega_{0}^4 Q^2}{2\pi v^4 \hbar} F N_{L},
\label{eq:Po1}
\end{align}
where $F$ is given by Eq.(\ref{eq:FTemu}). Finally $\mathcal{P}_{o} = \mathcal{P}_{o1}(L) - \mathcal{P}_{o1}(L\to e)$ gives Eq.(\ref{eq:Po}) in the main text.

\section{Acoustic phonon cooling in NLSMs}
\label{sec:acoustic_ph_nlsm}
Starting from Eq.(\ref{eq:pa1_pa2_0}) substitute $M_{\v{q}}$, $\omega_{\v{q}} = c q= c|\v{k}-\v{p}|$ and assume the lattice temperature is such that $k_B T_L \gg \hbar \omega_{\v{q}}$. This requires $T_L > T_{BG}$, as discussed in the main text. We obtain

\begin{align}
\mathcal{P}_{a1} &= \frac{2\pi  D^2 k_B T_L}{4 \rho V^2 c^2 } \sum_{n\v{k}} \sum_{m\v{p}} \omega_{\v{q}} ( f_{n\v{k}} - f_{m\v{p}}) \nn  \\ 
&~~~~~\times (1+s_{nm}\cos\theta) \delta(\epsilon_{n\v{k}} - \epsilon_{m\v{p}} - \hbar \omega_{\v{q}}) 
\label{eq:Pa1}
\end{align}
Now, we write the sums over momenta as integrals in cylindrical coordinates and define $k'=k-Q$. The integral over $k'$ is now from $-Q$ to $\infty$. The Fermi functions limit the size of $k'$ to values near $|k_F-Q|\ll Q$. Hence, we can extend the limits of integration of the radial component $k'$ to the whole real line with small error as long as $Q\gg |k_F-Q|,k_B T_e$.  Now transform $k',k_z$ variables to polar coordinates $\tilde{k},\phi_{\tilde{k}}$.  In terms of these variables, $k'=\tilde{k}\sin\phi_{\tilde{k}},k_z=\tilde{k}\cos\phi_{\tilde{k}}$ and the dispersion relation $\epsilon_{n\v{k}}$ is that of a Dirac cone in 2D. The energy difference becomes $\epsilon_{n\v{k}} - \epsilon_{m\v{p}} = v\hbar  (n\tilde{k} - m\tilde{p})$, and the delta function 

\begin{align}
\delta(\epsilon_{n\v{k}} - \epsilon_{m\v{p}} - \hbar \omega_{\v{q}}) &= \frac{1}{v\hbar}\delta(n \tilde{k} - m \tilde{p}- r q) \\
&= \frac{1}{v\hbar}\delta_{nm}\delta(\tilde{p}-\tilde{p}_0) [1 + O(r)], 
\label{eq:delta_p}
\end{align} 
fixes $\tilde{p}$ to $\tilde{p}_0= \tilde{k} - n r [q]_{\tilde{p}=\tilde{k}}$ (to first order in $r\equiv c/v$). $[q]_{\tilde{p}=\tilde{k}}$ is the phonon momentum to zero order, i.e., equal to $q=|\v{k}-\v{p}|$ with $\tilde{p}=\tilde{k}$. In the limit $Q\gg \tilde{k}_F$

\begin{align}
q^2 & \approx 2Q^2(1-\cos\phi) \\
1+s_{nm}\cos\theta &\approx 1+s_{nm}\cos\phi,
\end{align}
where $\phi \equiv \phi_{\v{k}} -\phi_{\v{p}}$. This means that the phonon momentum is effectively 2D in the plane $k_x$-$k_y$. Finally, we note $\hbar\omega_{\v{q}}$ and the Fermi function difference are both $O(r)$, and hence, we can set $\tilde{p}=\tilde{k}$ in the rest of the factors of Eq.~\ref{eq:Pa1}. Note that 

\begin{align}
 f_{n\tilde{k}}-f_{n\tilde{p}} = nr \tilde{q} \frac{\partial f_{n\tilde{k}}}{\partial \tilde{k}} + O(r^2),
\end{align}
is peaked at $\tilde{k} \sim \tilde{k}_F$. As a result Eq.~\ref{eq:Pa1} becomes 
 
\begin{align}
\mathcal{P}_{a1} &= -\frac{D^2 Q^4 k_B}{4\pi \rho v^2 \hbar } I_{1,+} T_L 
\label{eq:Pa1_v2}
\end{align}
where 

\begin{align}
I_{n\pm} \equiv \int_{0}^{\infty} \tilde{k}^n d\tilde{k}[ f_{c\tilde{k}} \pm (1-f_{v\tilde{k}}) ],
\label{eq:I_npm}
\end{align}
and $f_{n \tilde{k}} = [e^{(nv\hbar \tilde{k}-\mu)\beta}+1]^{-1}$

\section{Phonon cooling power in a Weyl node}
The quasiparticle energy near a Weyl node is $\epsilon_{n\v{k}} = nv\hbar k$ where $k=|\v{k}|$ is the magnitude of a 3D momentum and $n=\pm 1$ denotes the conduction (+1) and valence band (-1). The FS is a sphere in momentum space.

\subsection{Optical phonon relaxation}
Following the steps outlined in Appendix~\ref{sec:op_ph_nlsm} the cooling power of optical phonons in a Weyl node is 

\begin{align}
\mathcal{P}_{wo}=\frac{\omega_0^{6} g^2}{v^6 \hbar 2 \pi^3}H \left(N_{e} - N_{L} \right),
\end{align}
where 

\begin{align}
H &\equiv  \int_{-\infty}^{\infty} dx~ x^2 (x-1)^2 [f(\hbar\omega_{0}x -\hbar\omega_{0}) -f(\hbar\omega_{0}x) ] \nn \\
&= \frac{1}{30} + \bar{\mu}^4 + \frac{2\pi^2 \bar{\mu}^2}{\bar{\beta}^2} + \frac{7\pi^4}{15 \bar{\beta}^4},
\label{eq:H_exact}
\end{align}
has a simple closed form valid in all regimes of $\bar{\beta}\equiv \hbar\omega_{0}/k_B T_e$ and $\bar{\mu} \equiv \mu/\hbar\omega_{0}$.

\subsubsection{Low-density limit}
\textit{\textbf{case 1:} $\hbar\omega_0 \gg k_B T_e \gg T_L,\mu$}. If $\bar{\beta} \gg 1$ and $T_e \gg T_L$ then $N_e-N_L\sim e^{-\bar{\beta}}$. The energy and density 

\begin{align}
\mathcal{E} &= \frac{v\hbar I_{3+}}{2\pi^2}=  \frac{k_B^4 T_e^4 7\pi^2}{120 v^3 \hbar^3} + \frac{k_B^2 T_e^2 \mu^2}{4 v^3 \hbar^3} + \cdots \mu \ll k_B T_e 
\label{eq:Ew_and_n1_neutral} \\
n &= \frac{I_{2-}}{2\pi^2} =\frac{k_B^3 T_e^3 \mu}{6 v^3 \hbar^3} + \cdots ~~~~~~~~~~ \mu \ll k_B T_e,
\label{eq:Ew_and_n2_neutral}
\end{align}
together with Eq.(\ref{eq:H_exact}) and $d \mathcal{E}/dt = -\mathcal{P}_{wo}$ give
\begin{align}
\frac{d \bar{\beta}}{dt} = \tilde{\gamma}_{won1} (\bar{\beta}^5 + 14 \pi^4 \bar{\beta}) e^{-\bar{\beta}}.
\end{align} 
To leading order as $\bar{\beta}\gg 1$ the relaxation is an inverse log 
\begin{align}
k_{B} T_e = \frac{\hbar \omega_{0}}{\log(\gamma_{won1}\bar{\beta}_{0}^{5} t + e^{\bar{\beta}_0})},
\label{eq:relax_won1}
\end{align}
where $\bar{\beta}_0\equiv \hbar\omega_0/k_B T_0 \gg 1$, $\gamma_{won1}= g^2 \omega_{0}^{2}/14\pi^5 v^3 \hbar^2$, and the relaxation time scale is $e^{\bar{\beta}_0}/\gamma_{won1}\bar{\beta}_{0}^{5}$.\\

\textit{\textbf{case 2:} $k_B T_e \gg \hbar\omega_0,T_L,\mu$}. In this limit, $N_e- N_L \sim 1/\bar{\beta}$ and, hence, to leading order
\begin{align}
k_B T_e = \frac{\hbar \omega_0}{\gamma_{won2} t + \bar{\beta}_0}, 
\label{eq:relax_won2}
\end{align}
where $\gamma_{won2}= g^2 \omega_{0}^2/\pi^3 v^3 \hbar^2$.

\subsubsection{High-density limit}
\textit{\textbf{case 1:} $\mu,\hbar\omega_0 \gg k_B T_e \gg T_L$}. In the limit of large chemical potential $\mu \gg k_B T_e$ the energy and density are 

\begin{align}
\mathcal{E} &= \frac{1}{8\pi^2 v^3 \hbar^3} \bigg(\mu^4 + \frac{ 2\pi^2 \mu^2}{\beta^2} + \cdots\bigg)~~\mu \gg k_B T_e
\label{eq:Ew_and_n1_doped} \\ 
n &= \frac{1}{6\pi^2 v^3 \hbar^3} \bigg(\mu^3 + \frac{\pi^2 \mu}{\beta^2} + \cdots\bigg)~~~~\mu \gg k_B T_e.
\label{eq:Ew_and_n2_doped}
\end{align}
The last equation implies $\mu(T_e) = \epsilon_F - \pi^2 k_B^2 T_e^2/3\epsilon_F + \cdots$ and together with $N_e-N_L\sim e^{-\bar{\beta}}$ and Eq.(\ref{eq:H_exact}) give

\begin{align}
\frac{\partial \bar{\beta}}{\partial t} = \tilde{\gamma}_{wod1} \bar{\beta}^{3} \bigg(\frac{1}{30} + \bar{\mu}^4 + \frac{2\pi^2 \bar{\mu}^2}{\bar{\beta}^{2}} +\frac{7\pi^4}{15 \bar{\beta}^{4}} \bigg) e^{-\bar{\beta}}.
\end{align}
If $\bar{\mu}\bar{\beta} \gg \pi \sqrt{2}$, the leading behavior is given by the first two terms on the right and we obtain
\begin{align}
k_{B} T_e = \frac{\hbar \omega_{0}}{\log(\gamma_{wod1} \bar{\beta}^{3}_0 t + e^{\bar{\beta}_0})},
\label{eq:relax_wod1} 
\end{align}
with $\gamma_{wod1}= 3g^2 \omega_{0}^{4} (1/30 + \bar{\mu}^4)/\pi^2 v^3 \mu^2$ and time scale $e^{\bar{\beta}_0}/\gamma_{wod1} \bar{\beta}_0^{3}$. The time scale goes as $\sim n^{2/3}$  for $\bar{\mu}\ll 1$ and as $\sim n^{-2/3}$  for $\bar{\mu}\gg 1$.\\

\textit{\textbf{case 2:} $\mu\gg k_B T_e \gg \hbar\omega_0,T_L$}. In this case $N_e-N_L\sim 1/\bar{\beta}$ and from Eqs.(\ref{eq:H_exact}), (\ref{eq:Ew_and_n1_doped}), and (\ref{eq:Ew_and_n2_doped}) we obtain

\begin{align}
\frac{d \bar{\beta}}{dt} =  \tilde{\gamma}_{wod2}\bar{\beta}^2 \bigg( \frac{1}{30} + \bar{\mu}^4 + \frac{2\pi^2 \bar{\mu}^2}{\bar{\beta}^2} + \frac{7\pi^4}{15 \bar{\beta}^4} \bigg).
\end{align} 
If $\bar{\mu}\bar{\beta} \gg \pi \sqrt{2}$, the leading behavior is 
\begin{align}
\frac{d \bar{\beta}}{dt} =  \gamma_{wod2} \bar{\beta}^2,
\end{align} 
or
\begin{align}
k_{B} T_e = k_{B} T_0 - \hbar\omega_0 \gamma_{wod2}  t,
\label{eq:relax_wod2} 
\end{align}
where the relaxation rate $\gamma_{wod2}=3 g^2 \omega_{0}^{4}(1/30 + \bar{\mu}^4) /\pi^2 v^3 \mu^2$ goes as $\sim n^{-2/3}$ for $\bar{\mu}\ll 1$ and as $\sim n^{2/3}$ for $\bar{\mu} \gg 1$.

\subsection{Acoustic phonon relaxation}
The cooling power of acoustic phonons in a Weyl SM was studied in Ref.~\cite{Lundgren2015}. Here, we reproduce their results for comparison. Following the same steps as in Appendix~\ref{sec:acoustic_ph_nlsm}, we obtain

\begin{align}
\mathcal{P}_{wa1} &= -\frac{D^2 k_{B}}{\hbar \rho v^2 \pi^3}  I_{5+} T_L 
\end{align}
and, hence, the cooling power of acoustic phonons is~\cite{Lundgren2015} 

\begin{align}
\mathcal{P}_{wa}&= \mathcal{P}_{wa1}(L) - \mathcal{P}_{wa1}(L\to e)\nn\\
&= \frac{D^2 k_{B}}{ \rho v^2 \hbar \pi^3} I_{5+} (T_e-T_L) 
\label{eq:pwa}
\end{align}
where 

\begin{align}
I_{n\pm} \equiv \int_{0}^{\infty} k^n d k ~[ f_{ck} \pm (1-f_{vk} ) ],
\label{eq:Ipm_def_weyl}
\end{align}
and $f_{nk}= [e^{\beta(nv\hbar k - \mu)}+1]^{-1}$.

\subsubsection{Low-density limit}
If $\mu=0$ we can integrate (\ref{eq:pwa}) analytically to obtain

\begin{align}
\mathcal{P}_{wa}= \frac{D^2 k_B^7 T_e^6 31 \pi^3}{\rho v^{8}\hbar^7 126} (T_e-T_L).
\end{align}
The energy and density of a Weyl node are given by Eqs.(\ref{eq:Ew_and_n1_neutral}) and (\ref{eq:Ew_and_n2_neutral}) and together with $d \mathcal{E}/dt = -\mathcal{P}_{wa}$ we obtain
\begin{align}
\frac{dT_e}{dt} = -\gamma_{wan}T_e^3 (T_e-T_L) 
\label{eq:weyl_acoustic_neutral}
\end{align}
where  $\gamma_{wan}= 155\pi D^2 k_B^3/147 \rho v^5 \hbar^4$. If $T_e \gg T_L$, $T_e$ relaxes as a power law~\cite{Lundgren2015}
\begin{align}
T_e = \frac{T_0}{(1 + 3\gamma_{wan} T_0^3 t )^{1/3}},
\end{align}
with time scale $1/3\gamma_{wan}T_0^3$, but if $T_e \gtrsim T_L$, $T_e$ relaxes exponentially with time scale $1/\gamma_{wan}T_L^3$

\subsubsection{High-density limit}
Using Eqs.(\ref{eq:Ew_and_n1_doped}), (\ref{eq:Ew_and_n2_doped}) and (\ref{eq:pwa}) we have to leading order~\cite{Lundgren2015}

\begin{align}
\frac{d }{dt}k_B T_e  \sim -\gamma_{wad}\frac{1}{T_e} (T_e-T_L),
\label{eq:relax_wad}
\end{align}
where $\gamma_{wad} = D^2 \mu^4/ 3\pi^3 \rho v^5 \hbar^4 $. So, for $T_e\gg T_L$ the relaxation is linear 

\begin{align}
k_B T_e \sim k_B T_0 - \gamma_{wad}t,
\end{align}
with a rate that goes as $\sim n^{4/3}$ with electron density.  In the limit $T_e \gtrsim T_L$, the relaxation is exponential with time scale $T_L/\gamma_{wad} \sim n^{-4/3}$.

\section{Phonon cooling power in graphene} 
In this Appendix, we calculate the temperature relaxation in graphene due to acoustic and optical phonons. The dispersion relation of quasiparticles near a nodal point is $\epsilon_{n \v{k}} = nv\hbar k$, where $k=|\v{k}|$ is a 2D wave vector.

\subsection{Optical phonon relaxation}
Following the same steps as in Appendix~\ref{sec:op_ph_nlsm}, and using $\omega_{\v{q}}= \omega_0$ and $M_{\v{q}}=g^2/V$, we obtain~\cite{Bistritzer2009}

\begin{align}
\mathcal{P}_{go}= \frac{g^2\omega^4_{0}}{2\pi v^4 \hbar}F(\mu,T_e) (N_e - N_L)
\label{eq:Pgo}
\end{align}
where $F$ is given in Eq.(\ref{eq:FTemu}), and Eqs.(\ref{eq:Fu_exp}) and (\ref{eq:Fon2}) are valid for graphene too.

\subsubsection{Low-density limit}
\textit{\textbf{case 1:} $\hbar\omega_0 \gg k_B T_e \gg T_L,\mu$}. The energy and particle density $\mathcal{E} = v\hbar I_{2+}/2\pi$ and $n = I_{1-}/2\pi$ are the same as for NLSMs [Eqs.(\ref{eq:E_and_n1}) and (\ref{eq:E_and_n2})]  except for a trivial factor of $Q$, which is absent in graphene. Defining as before $\bar{\beta}\equiv \hbar\omega_{0}/k_B T_e$ and using (\ref{eq:Pgo}) we find 
\begin{align}
\frac{\partial \bar{\beta}}{\partial t} = \gamma_{gon1} \big( \bar{\beta}^{4} + 36\zeta{(3)} \bar{\beta} + \cdots\big) e^{-\bar{\beta}}
\end{align}
where $\gamma_{gon1} = g^2\omega_0/54\zeta{(3)}v^2 \hbar^2$ and to leading order in $\bar{\beta}\gg 1$ we obtain an inverse log form
\begin{align}
k_{B} T_e = \frac{\hbar \omega_{0}}{\log(\gamma_{gon1} \bar{\beta}^{4}_0 t + e^{\bar{\beta}_0})}, 
\label{eq:relax_gon1}
\end{align}

\textit{\textbf{case 2:} $k_B T_e\gg \hbar \omega_{0},  k_B T_L,\mu$}. In the opposite regime where $k_B T_e\gg \hbar \omega_{0}$ the same considerations apply as in NLSMs and we obtain an exponential relaxation [see Eq.(\ref{eq:Pon2})], but the factor of $Q$ is absent in graphene.

\subsubsection{High-density limit}
The considerations of Sec.~\ref{sec:nlsm_op_doped} apply for graphene as well. We obtain the same relaxation forms as in Eq.(\ref{eq:nlsm_op_doped1}) and (\ref{eq:relax_nlsm_over_case2}) but the factor of $Q$ is absent in graphene.

\subsection{Acoustic phonon relaxation}
The temperature relaxation due to acoustic phonon scattering in NLSMs and graphene are different because the nodal line affects NLSMs in a nontrivial way (see Appendix~\ref{sec:acoustic_ph_nlsm}). The temperature relaxation in graphene due to acoustic phonons was presented by Bistritzer and MacDonald~\cite{Bistritzer2009} and is included here for comparison with NLSMs. From Eq.~\ref{eq:P_gen} we obtain 
\begin{align}
\mathcal{P}_{ga} = \frac{D^2 k_B}{\rho v^2 \hbar 2\pi} I_{3+} (T_e - T_L)  
\end{align}
where $I_{3+}$ is given in Eq.(\ref{eq:I_npm}) and in closed form in Ref.~\cite{Viljas2010}.

\subsubsection{Low-density limit}
Eqs.(\ref{eq:E_and_n1}) and (\ref{eq:E_and_n2}) in the limit $\mu \ll k_B T_e$ are also valid for graphene if we omit the factor of $Q$. From the equation of motion $(\partial \mathcal{E}/\partial T_e)(\partial T_e/\partial t) = -\mathcal{P}_{ga}$ we obtain

\begin{align}
\frac{d T_e}{d t} = -\gamma_{gan}T_e^2 (T_e -T_L)
\label{eq:graphene_acoustic_neutral}
\end{align}
where $\gamma_{gan} = D^2 k_B^{2} 7 \pi^4/540\zeta{(3)}\rho v^4 \hbar^3$. This means $T_e = T_0/(1+t/\tau_{gan})^{1/2}$ for $T_e \gg T_L$. $T_e$ is an exponential function with time scale $1/\gamma_{gan} T_L^2$ for $T_e\gtrsim T_L$.

\subsubsection{High-density limit}
Eqs.(\ref{eq:E_largemu}) and (\ref{eq:n_largemu}) are also valid for graphene after omitting the factor of $Q$ and we obtain 
\begin{align}
\frac{d T_e}{d t} = -\gamma_{gad}\frac{1}{T_e} (T_e -T_L),
\label{eq:relax_gad}
\end{align}
where $\gamma_{gad} = 3D^2 \mu^3/4\pi^2 k_B \rho v^4 \hbar^3$. This means the relaxation is linear for $T_e\gg T_L$ and the rate scales as $\sim n^{3/2}$ with electron density. The relaxation is exponential for $T_e \gtrsim T_L$.


\begin{thebibliography}{42}%
\makeatletter
\providecommand \@ifxundefined [1]{%
 \@ifx{#1\undefined}
}%
\providecommand \@ifnum [1]{%
 \ifnum #1\expandafter \@firstoftwo
 \else \expandafter \@secondoftwo
 \fi
}%
\providecommand \@ifx [1]{%
 \ifx #1\expandafter \@firstoftwo
 \else \expandafter \@secondoftwo
 \fi
}%
\providecommand \natexlab [1]{#1}%
\providecommand \enquote  [1]{``#1''}%
\providecommand \bibnamefont  [1]{#1}%
\providecommand \bibfnamefont [1]{#1}%
\providecommand \citenamefont [1]{#1}%
\providecommand \href@noop [0]{\@secondoftwo}%
\providecommand \href [0]{\begingroup \@sanitize@url \@href}%
\providecommand \@href[1]{\@@startlink{#1}\@@href}%
\providecommand \@@href[1]{\endgroup#1\@@endlink}%
\providecommand \@sanitize@url [0]{\catcode `\\12\catcode `\$12\catcode
  `\&12\catcode `\#12\catcode `\^12\catcode `\_12\catcode `\%12\relax}%
\providecommand \@@startlink[1]{}%
\providecommand \@@endlink[0]{}%
\providecommand \url  [0]{\begingroup\@sanitize@url \@url }%
\providecommand \@url [1]{\endgroup\@href {#1}{\urlprefix }}%
\providecommand \urlprefix  [0]{URL }%
\providecommand \Eprint [0]{\href }%
\providecommand \doibase [0]{https://doi.org/}%
\providecommand \selectlanguage [0]{\@gobble}%
\providecommand \bibinfo  [0]{\@secondoftwo}%
\providecommand \bibfield  [0]{\@secondoftwo}%
\providecommand \translation [1]{[#1]}%
\providecommand \BibitemOpen [0]{}%
\providecommand \bibitemStop [0]{}%
\providecommand \bibitemNoStop [0]{.\EOS\space}%
\providecommand \EOS [0]{\spacefactor3000\relax}%
\providecommand \BibitemShut  [1]{\csname bibitem#1\endcsname}%
\let\auto@bib@innerbib\@empty
\bibitem [{\citenamefont {Gantmakher}\ and\ \citenamefont
  {Levinson}(1987)}]{Gantmakher1987}%
  \BibitemOpen
  \bibfield  {author} {\bibinfo {author} {\bibfnamefont {V.~F.}\ \bibnamefont
  {Gantmakher}}\ and\ \bibinfo {author} {\bibfnamefont {Y.~B.}\ \bibnamefont
  {Levinson}},\ }\href@noop {} {\emph {\bibinfo {title} {Carrier scattering in
  metals and semiconductors}}}\ (\bibinfo  {publisher} {Elsevier Science
  Publishers, NY},\ \bibinfo {year} {1987})\BibitemShut {NoStop}%
\bibitem [{\citenamefont {Sze}\ \emph {et~al.}(2021)\citenamefont {Sze},
  \citenamefont {Li},\ and\ \citenamefont {Ng}}]{Sze2021}%
  \BibitemOpen
  \bibfield  {author} {\bibinfo {author} {\bibfnamefont {S.~M.}\ \bibnamefont
  {Sze}}, \bibinfo {author} {\bibfnamefont {Y.}~\bibnamefont {Li}},\ and\
  \bibinfo {author} {\bibfnamefont {K.~K.}\ \bibnamefont {Ng}},\ }\href@noop {}
  {\emph {\bibinfo {title} {Physics of Semiconductor Devices}}}\ (\bibinfo
  {publisher} {Wiley; 4th edition},\ \bibinfo {year} {2021})\BibitemShut
  {NoStop}%
\bibitem [{\citenamefont {Allen}(1987)}]{Allen1987}%
  \BibitemOpen
  \bibfield  {author} {\bibinfo {author} {\bibfnamefont {P.~B.}\ \bibnamefont
  {Allen}},\ }\bibfield  {title} {\bibinfo {title} {Theory of thermal
  relaxation of electrons in metals},\ }\href@noop {} {\bibfield  {journal}
  {\bibinfo  {journal} {Phys. Rev. Lett.}\ }\textbf {\bibinfo {volume} {59}},\
  \bibinfo {pages} {1460} (\bibinfo {year} {1987})}\BibitemShut {NoStop}%
\bibitem [{\citenamefont {Wellstood}\ \emph {et~al.}(1994)\citenamefont
  {Wellstood}, \citenamefont {Urbina},\ and\ \citenamefont
  {Clarke}}]{Wellstood1994}%
  \BibitemOpen
  \bibfield  {author} {\bibinfo {author} {\bibfnamefont {F.~C.}\ \bibnamefont
  {Wellstood}}, \bibinfo {author} {\bibfnamefont {C.}~\bibnamefont {Urbina}},\
  and\ \bibinfo {author} {\bibfnamefont {J.}~\bibnamefont {Clarke}},\
  }\bibfield  {title} {\bibinfo {title} {Hot-electron effects in metals},\
  }\href@noop {} {\bibfield  {journal} {\bibinfo  {journal} {Phys. Rev. B}\
  }\textbf {\bibinfo {volume} {49}},\ \bibinfo {pages} {5942} (\bibinfo {year}
  {1994})}\BibitemShut {NoStop}%
\bibitem [{\citenamefont {Massicotte}\ \emph {et~al.}(2021)\citenamefont
  {Massicotte}, \citenamefont {Soavi}, \citenamefont {Principi},\ and\
  \citenamefont {Tielrooij}}]{Massicotte2021}%
  \BibitemOpen
  \bibfield  {author} {\bibinfo {author} {\bibfnamefont {M.}~\bibnamefont
  {Massicotte}}, \bibinfo {author} {\bibfnamefont {G.}~\bibnamefont {Soavi}},
  \bibinfo {author} {\bibfnamefont {A.}~\bibnamefont {Principi}},\ and\
  \bibinfo {author} {\bibfnamefont {K.-J.}\ \bibnamefont {Tielrooij}},\
  }\bibfield  {title} {\bibinfo {title} {Hot carriers in graphene –
  fundamentals and applications},\ }\href@noop {} {\bibfield  {journal}
  {\bibinfo  {journal} {Nanoscale}\ }\textbf {\bibinfo {volume} {13}},\
  \bibinfo {pages} {8376} (\bibinfo {year} {2021})}\BibitemShut {NoStop}%
\bibitem [{\citenamefont {Sergeev}\ and\ \citenamefont
  {Mitin}(2000)}]{Sergeev2000}%
  \BibitemOpen
  \bibfield  {author} {\bibinfo {author} {\bibfnamefont {A.}~\bibnamefont
  {Sergeev}}\ and\ \bibinfo {author} {\bibfnamefont {V.}~\bibnamefont
  {Mitin}},\ }\bibfield  {title} {\bibinfo {title} {Electron-phonon interaction
  in disordered conductors: Static and vibrating scattering potentials},\
  }\href@noop {} {\bibfield  {journal} {\bibinfo  {journal} {Phys. Rev. B}\
  }\textbf {\bibinfo {volume} {61}},\ \bibinfo {pages} {6041} (\bibinfo {year}
  {2000})}\BibitemShut {NoStop}%
\bibitem [{\citenamefont {Karvonen}\ \emph {et~al.}(2005)\citenamefont
  {Karvonen}, \citenamefont {Taskinen},\ and\ \citenamefont
  {Maasilta}}]{Karvonen2005}%
  \BibitemOpen
  \bibfield  {author} {\bibinfo {author} {\bibfnamefont {J.~T.}\ \bibnamefont
  {Karvonen}}, \bibinfo {author} {\bibfnamefont {L.~J.}\ \bibnamefont
  {Taskinen}},\ and\ \bibinfo {author} {\bibfnamefont {I.~J.}\ \bibnamefont
  {Maasilta}},\ }\bibfield  {title} {\bibinfo {title} {Observation of
  disorder-induced weakening of electron-phonon interaction in thin noble-metal
  films},\ }\href@noop {} {\bibfield  {journal} {\bibinfo  {journal} {Phys.
  Rev. B}\ }\textbf {\bibinfo {volume} {72}},\ \bibinfo {pages} {012302}
  (\bibinfo {year} {2005})}\BibitemShut {NoStop}%
\bibitem [{\citenamefont {Hekking}\ \emph {et~al.}(2008)\citenamefont
  {Hekking}, \citenamefont {Niskanen},\ and\ \citenamefont
  {Pekola}}]{Hekking2008}%
  \BibitemOpen
  \bibfield  {author} {\bibinfo {author} {\bibfnamefont {F.~W.~J.}\
  \bibnamefont {Hekking}}, \bibinfo {author} {\bibfnamefont {A.~O.}\
  \bibnamefont {Niskanen}},\ and\ \bibinfo {author} {\bibfnamefont {J.~P.}\
  \bibnamefont {Pekola}},\ }\bibfield  {title} {\bibinfo {title}
  {Electron-phonon coupling and longitudinal mechanical-mode cooling in a
  metallic nanowire},\ }\href@noop {} {\bibfield  {journal} {\bibinfo
  {journal} {Phys. Rev. B}\ }\textbf {\bibinfo {volume} {77}},\ \bibinfo
  {pages} {033401} (\bibinfo {year} {2008})}\BibitemShut {NoStop}%
\bibitem [{\citenamefont {Kubakaddi}(2009)}]{Kubakaddi2009}%
  \BibitemOpen
  \bibfield  {author} {\bibinfo {author} {\bibfnamefont {S.~S.}\ \bibnamefont
  {Kubakaddi}},\ }\bibfield  {title} {\bibinfo {title} {Interaction of massless
  dirac electrons with acoustic phonons in graphene at low temperatures},\
  }\href@noop {} {\bibfield  {journal} {\bibinfo  {journal} {Phys. Rev. B}\
  }\textbf {\bibinfo {volume} {79}},\ \bibinfo {pages} {075417} (\bibinfo
  {year} {2009})}\BibitemShut {NoStop}%
\bibitem [{\citenamefont {Armitage}\ \emph {et~al.}(2018)\citenamefont
  {Armitage}, \citenamefont {Mele},\ and\ \citenamefont
  {Vishwanath}}]{Armitage2018}%
  \BibitemOpen
  \bibfield  {author} {\bibinfo {author} {\bibfnamefont {N.~P.}\ \bibnamefont
  {Armitage}}, \bibinfo {author} {\bibfnamefont {E.~J.}\ \bibnamefont {Mele}},\
  and\ \bibinfo {author} {\bibfnamefont {A.}~\bibnamefont {Vishwanath}},\
  }\bibfield  {title} {\bibinfo {title} {Weyl and dirac semimetals in
  three-dimensional solids},\ }\href@noop {} {\bibfield  {journal} {\bibinfo
  {journal} {Rev. Mod. Phys.}\ }\textbf {\bibinfo {volume} {90}},\ \bibinfo
  {pages} {015001} (\bibinfo {year} {2018})}\BibitemShut {NoStop}%
\bibitem [{\citenamefont {Stormer}\ \emph {et~al.}(1990)\citenamefont
  {Stormer}, \citenamefont {Pfeiffer}, \citenamefont {Baldwin},\ and\
  \citenamefont {West}}]{Stormer1990}%
  \BibitemOpen
  \bibfield  {author} {\bibinfo {author} {\bibfnamefont {H.~L.}\ \bibnamefont
  {Stormer}}, \bibinfo {author} {\bibfnamefont {L.~N.}\ \bibnamefont
  {Pfeiffer}}, \bibinfo {author} {\bibfnamefont {K.~W.}\ \bibnamefont
  {Baldwin}},\ and\ \bibinfo {author} {\bibfnamefont {K.~W.}\ \bibnamefont
  {West}},\ }\bibfield  {title} {\bibinfo {title} {Observation of a
  bloch-gr\"uneisen regime in two-dimensional electron transport},\ }\href@noop
  {} {\bibfield  {journal} {\bibinfo  {journal} {Phys. Rev. B}\ }\textbf
  {\bibinfo {volume} {41}},\ \bibinfo {pages} {1278} (\bibinfo {year}
  {1990})}\BibitemShut {NoStop}%
\bibitem [{\citenamefont {Efetov}\ and\ \citenamefont
  {Kim}(2010)}]{Efetov2010}%
  \BibitemOpen
  \bibfield  {author} {\bibinfo {author} {\bibfnamefont {D.~K.}\ \bibnamefont
  {Efetov}}\ and\ \bibinfo {author} {\bibfnamefont {P.}~\bibnamefont {Kim}},\
  }\bibfield  {title} {\bibinfo {title} {Controlling electron-phonon
  interactions in graphene at ultrahigh carrier densities},\ }\href@noop {}
  {\bibfield  {journal} {\bibinfo  {journal} {Phys. Rev. Lett.}\ }\textbf
  {\bibinfo {volume} {105}},\ \bibinfo {pages} {256805} (\bibinfo {year}
  {2010})}\BibitemShut {NoStop}%
\bibitem [{\citenamefont {Tse}\ and\ \citenamefont
  {Das~Sarma}(2009)}]{Tse2009}%
  \BibitemOpen
  \bibfield  {author} {\bibinfo {author} {\bibfnamefont {W.-K.}\ \bibnamefont
  {Tse}}\ and\ \bibinfo {author} {\bibfnamefont {S.}~\bibnamefont
  {Das~Sarma}},\ }\bibfield  {title} {\bibinfo {title} {Energy relaxation of
  hot dirac fermions in graphene},\ }\href@noop {} {\bibfield  {journal}
  {\bibinfo  {journal} {Phys. Rev. B}\ }\textbf {\bibinfo {volume} {79}},\
  \bibinfo {pages} {235406} (\bibinfo {year} {2009})}\BibitemShut {NoStop}%
\bibitem [{\citenamefont {Song}\ \emph {et~al.}(2012)\citenamefont {Song},
  \citenamefont {Reizer},\ and\ \citenamefont {Levitov}}]{Song2012}%
  \BibitemOpen
  \bibfield  {author} {\bibinfo {author} {\bibfnamefont {J.~C.~W.}\
  \bibnamefont {Song}}, \bibinfo {author} {\bibfnamefont {M.~Y.}\ \bibnamefont
  {Reizer}},\ and\ \bibinfo {author} {\bibfnamefont {L.~S.}\ \bibnamefont
  {Levitov}},\ }\bibfield  {title} {\bibinfo {title} {Disorder-assisted
  electron-phonon scattering and cooling pathways in graphene},\ }\href@noop {}
  {\bibfield  {journal} {\bibinfo  {journal} {Phys. Rev. Lett.}\ }\textbf
  {\bibinfo {volume} {109}},\ \bibinfo {pages} {106602} (\bibinfo {year}
  {2012})}\BibitemShut {NoStop}%
\bibitem [{\citenamefont {Graham}\ \emph {et~al.}(2013)\citenamefont {Graham},
  \citenamefont {Shi}, \citenamefont {Ralph}, \citenamefont {Park},\ and\
  \citenamefont {McEuen}}]{Graham2013}%
  \BibitemOpen
  \bibfield  {author} {\bibinfo {author} {\bibfnamefont {M.~W.}\ \bibnamefont
  {Graham}}, \bibinfo {author} {\bibfnamefont {S.-F.}\ \bibnamefont {Shi}},
  \bibinfo {author} {\bibfnamefont {D.~C.}\ \bibnamefont {Ralph}}, \bibinfo
  {author} {\bibfnamefont {J.}~\bibnamefont {Park}},\ and\ \bibinfo {author}
  {\bibfnamefont {P.~L.}\ \bibnamefont {McEuen}},\ }\bibfield  {title}
  {\bibinfo {title} {Photocurrent measurements of supercollision cooling in
  graphene},\ }\href@noop {} {\bibfield  {journal} {\bibinfo  {journal} {Nature
  Physics}\ }\textbf {\bibinfo {volume} {9}},\ \bibinfo {pages} {103} (\bibinfo
  {year} {2013})}\BibitemShut {NoStop}%
\bibitem [{\citenamefont {Bistritzer}\ and\ \citenamefont
  {MacDonald}(2009)}]{Bistritzer2009}%
  \BibitemOpen
  \bibfield  {author} {\bibinfo {author} {\bibfnamefont {R.}~\bibnamefont
  {Bistritzer}}\ and\ \bibinfo {author} {\bibfnamefont {A.~H.}\ \bibnamefont
  {MacDonald}},\ }\bibfield  {title} {\bibinfo {title} {Electronic cooling in
  graphene},\ }\href@noop {} {\bibfield  {journal} {\bibinfo  {journal} {Phys.
  Rev. Lett.}\ }\textbf {\bibinfo {volume} {102}},\ \bibinfo {pages} {206410}
  (\bibinfo {year} {2009})}\BibitemShut {NoStop}%
\bibitem [{\citenamefont {Viljas}\ and\ \citenamefont
  {Heikkil\"a}(2010)}]{Viljas2010}%
  \BibitemOpen
  \bibfield  {author} {\bibinfo {author} {\bibfnamefont {J.~K.}\ \bibnamefont
  {Viljas}}\ and\ \bibinfo {author} {\bibfnamefont {T.~T.}\ \bibnamefont
  {Heikkil\"a}},\ }\bibfield  {title} {\bibinfo {title} {Electron-phonon heat
  transfer in monolayer and bilayer graphene},\ }\href@noop {} {\bibfield
  {journal} {\bibinfo  {journal} {Phys. Rev. B}\ }\textbf {\bibinfo {volume}
  {81}},\ \bibinfo {pages} {245404} (\bibinfo {year} {2010})}\BibitemShut
  {NoStop}%
\bibitem [{\citenamefont {Wang}\ \emph {et~al.}(2012)\citenamefont {Wang},
  \citenamefont {Hsieh}, \citenamefont {Sie}, \citenamefont {Steinberg},
  \citenamefont {Gardner}, \citenamefont {Lee}, \citenamefont
  {Jarillo-Herrero},\ and\ \citenamefont {Gedik}}]{Wang2012}%
  \BibitemOpen
  \bibfield  {author} {\bibinfo {author} {\bibfnamefont {Y.~H.}\ \bibnamefont
  {Wang}}, \bibinfo {author} {\bibfnamefont {D.}~\bibnamefont {Hsieh}},
  \bibinfo {author} {\bibfnamefont {E.~J.}\ \bibnamefont {Sie}}, \bibinfo
  {author} {\bibfnamefont {H.}~\bibnamefont {Steinberg}}, \bibinfo {author}
  {\bibfnamefont {D.~R.}\ \bibnamefont {Gardner}}, \bibinfo {author}
  {\bibfnamefont {Y.~S.}\ \bibnamefont {Lee}}, \bibinfo {author} {\bibfnamefont
  {P.}~\bibnamefont {Jarillo-Herrero}},\ and\ \bibinfo {author} {\bibfnamefont
  {N.}~\bibnamefont {Gedik}},\ }\bibfield  {title} {\bibinfo {title}
  {Measurement of intrinsic dirac fermion cooling on the surface of the
  topological insulator ${\mathrm{bi}}_{2}{\mathrm{se}}_{3}$ using
  time-resolved and angle-resolved photoemission spectroscopy},\ }\href@noop {}
  {\bibfield  {journal} {\bibinfo  {journal} {Phys. Rev. Lett.}\ }\textbf
  {\bibinfo {volume} {109}},\ \bibinfo {pages} {127401} (\bibinfo {year}
  {2012})}\BibitemShut {NoStop}%
\bibitem [{\citenamefont {Coulter}\ \emph {et~al.}(2019)\citenamefont
  {Coulter}, \citenamefont {Osterhoudt}, \citenamefont {Garcia}, \citenamefont
  {Wang}, \citenamefont {Plisson}, \citenamefont {Shen}, \citenamefont {Ni},
  \citenamefont {Burch},\ and\ \citenamefont {Narang}}]{Coulter2019}%
  \BibitemOpen
  \bibfield  {author} {\bibinfo {author} {\bibfnamefont {J.}~\bibnamefont
  {Coulter}}, \bibinfo {author} {\bibfnamefont {G.~B.}\ \bibnamefont
  {Osterhoudt}}, \bibinfo {author} {\bibfnamefont {C.~A.~C.}\ \bibnamefont
  {Garcia}}, \bibinfo {author} {\bibfnamefont {Y.}~\bibnamefont {Wang}},
  \bibinfo {author} {\bibfnamefont {V.~M.}\ \bibnamefont {Plisson}}, \bibinfo
  {author} {\bibfnamefont {B.}~\bibnamefont {Shen}}, \bibinfo {author}
  {\bibfnamefont {N.}~\bibnamefont {Ni}}, \bibinfo {author} {\bibfnamefont
  {K.~S.}\ \bibnamefont {Burch}},\ and\ \bibinfo {author} {\bibfnamefont
  {P.}~\bibnamefont {Narang}},\ }\bibfield  {title} {\bibinfo {title}
  {Uncovering electron-phonon scattering and phonon dynamics in type-i weyl
  semimetals},\ }\href@noop {} {\bibfield  {journal} {\bibinfo  {journal}
  {Phys. Rev. B}\ }\textbf {\bibinfo {volume} {100}},\ \bibinfo {pages}
  {220301(R)} (\bibinfo {year} {2019})}\BibitemShut {NoStop}%
\bibitem [{\citenamefont {Osterhoudt}\ \emph {et~al.}(2021)\citenamefont
  {Osterhoudt}, \citenamefont {Wang}, \citenamefont {Garcia}, \citenamefont
  {Plisson}, \citenamefont {Gooth}, \citenamefont {Felser}, \citenamefont
  {Narang},\ and\ \citenamefont {Burch}}]{Osterhoudt2021}%
  \BibitemOpen
  \bibfield  {author} {\bibinfo {author} {\bibfnamefont {G.~B.}\ \bibnamefont
  {Osterhoudt}}, \bibinfo {author} {\bibfnamefont {Y.}~\bibnamefont {Wang}},
  \bibinfo {author} {\bibfnamefont {C.~A.~C.}\ \bibnamefont {Garcia}}, \bibinfo
  {author} {\bibfnamefont {V.~M.}\ \bibnamefont {Plisson}}, \bibinfo {author}
  {\bibfnamefont {J.}~\bibnamefont {Gooth}}, \bibinfo {author} {\bibfnamefont
  {C.}~\bibnamefont {Felser}}, \bibinfo {author} {\bibfnamefont
  {P.}~\bibnamefont {Narang}},\ and\ \bibinfo {author} {\bibfnamefont {K.~S.}\
  \bibnamefont {Burch}},\ }\bibfield  {title} {\bibinfo {title} {Evidence for
  dominant phonon-electron scattering in weyl semimetal ${\mathrm{wp}}_{2}$},\
  }\href@noop {} {\bibfield  {journal} {\bibinfo  {journal} {Phys. Rev. X}\
  }\textbf {\bibinfo {volume} {11}},\ \bibinfo {pages} {011017} (\bibinfo
  {year} {2021})}\BibitemShut {NoStop}%
\bibitem [{\citenamefont {Lundgren}\ and\ \citenamefont
  {Fiete}(2015)}]{Lundgren2015}%
  \BibitemOpen
  \bibfield  {author} {\bibinfo {author} {\bibfnamefont {R.}~\bibnamefont
  {Lundgren}}\ and\ \bibinfo {author} {\bibfnamefont {G.~A.}\ \bibnamefont
  {Fiete}},\ }\bibfield  {title} {\bibinfo {title} {Electronic cooling in weyl
  and dirac semimetals},\ }\href@noop {} {\bibfield  {journal} {\bibinfo
  {journal} {Phys. Rev. B}\ }\textbf {\bibinfo {volume} {92}},\ \bibinfo
  {pages} {125139} (\bibinfo {year} {2015})}\BibitemShut {NoStop}%
\bibitem [{\citenamefont {Burkov}\ \emph {et~al.}(2011)\citenamefont {Burkov},
  \citenamefont {Hook},\ and\ \citenamefont {Balents}}]{Burkov2011}%
  \BibitemOpen
  \bibfield  {author} {\bibinfo {author} {\bibfnamefont {A.~A.}\ \bibnamefont
  {Burkov}}, \bibinfo {author} {\bibfnamefont {M.~D.}\ \bibnamefont {Hook}},\
  and\ \bibinfo {author} {\bibfnamefont {L.}~\bibnamefont {Balents}},\
  }\bibfield  {title} {\bibinfo {title} {Topological nodal semimetals},\
  }\href@noop {} {\bibfield  {journal} {\bibinfo  {journal} {Phys. Rev. B}\
  }\textbf {\bibinfo {volume} {84}},\ \bibinfo {pages} {235126} (\bibinfo
  {year} {2011})}\BibitemShut {NoStop}%
\bibitem [{\citenamefont {Kim}\ \emph {et~al.}(2015)\citenamefont {Kim},
  \citenamefont {Wieder}, \citenamefont {Kane},\ and\ \citenamefont
  {Rappe}}]{Kim2015}%
  \BibitemOpen
  \bibfield  {author} {\bibinfo {author} {\bibfnamefont {Y.}~\bibnamefont
  {Kim}}, \bibinfo {author} {\bibfnamefont {B.~J.}\ \bibnamefont {Wieder}},
  \bibinfo {author} {\bibfnamefont {C.~L.}\ \bibnamefont {Kane}},\ and\
  \bibinfo {author} {\bibfnamefont {A.~M.}\ \bibnamefont {Rappe}},\ }\bibfield
  {title} {\bibinfo {title} {Dirac line nodes in inversion-symmetric
  crystals},\ }\href@noop {} {\bibfield  {journal} {\bibinfo  {journal} {Phys.
  Rev. Lett.}\ }\textbf {\bibinfo {volume} {115}},\ \bibinfo {pages} {036806}
  (\bibinfo {year} {2015})}\BibitemShut {NoStop}%
\bibitem [{\citenamefont {Fang}\ \emph {et~al.}(2015)\citenamefont {Fang},
  \citenamefont {Chen}, \citenamefont {Kee},\ and\ \citenamefont
  {Fu}}]{Fang2015}%
  \BibitemOpen
  \bibfield  {author} {\bibinfo {author} {\bibfnamefont {C.}~\bibnamefont
  {Fang}}, \bibinfo {author} {\bibfnamefont {Y.}~\bibnamefont {Chen}}, \bibinfo
  {author} {\bibfnamefont {H.-Y.}\ \bibnamefont {Kee}},\ and\ \bibinfo {author}
  {\bibfnamefont {L.}~\bibnamefont {Fu}},\ }\bibfield  {title} {\bibinfo
  {title} {Topological nodal line semimetals with and without spin-orbital
  coupling},\ }\href@noop {} {\bibfield  {journal} {\bibinfo  {journal} {Phys.
  Rev. B}\ }\textbf {\bibinfo {volume} {92}},\ \bibinfo {pages} {081201(R)}
  (\bibinfo {year} {2015})}\BibitemShut {NoStop}%
\bibitem [{\citenamefont {Fang}\ \emph {et~al.}(2016)\citenamefont {Fang},
  \citenamefont {Weng}, \citenamefont {Dai},\ and\ \citenamefont
  {Fang}}]{Fang2016}%
  \BibitemOpen
  \bibfield  {author} {\bibinfo {author} {\bibfnamefont {C.}~\bibnamefont
  {Fang}}, \bibinfo {author} {\bibfnamefont {H.}~\bibnamefont {Weng}}, \bibinfo
  {author} {\bibfnamefont {X.}~\bibnamefont {Dai}},\ and\ \bibinfo {author}
  {\bibfnamefont {Z.}~\bibnamefont {Fang}},\ }\bibfield  {title} {\bibinfo
  {title} {Topological nodal line semimetals},\ }\href@noop {} {\bibfield
  {journal} {\bibinfo  {journal} {Chinese Physics B}\ }\textbf {\bibinfo
  {volume} {25}},\ \bibinfo {pages} {117106} (\bibinfo {year}
  {2016})}\BibitemShut {NoStop}%
\bibitem [{\citenamefont {Bian}\ \emph {et~al.}(2016)\citenamefont {Bian},
  \citenamefont {Chang}, \citenamefont {Sankar}, \citenamefont {Xu},
  \citenamefont {Zheng}, \citenamefont {Neupert}, \citenamefont {Chiu},
  \citenamefont {Huang}, \citenamefont {Chang}, \citenamefont {Belopolski},
  \citenamefont {Sanchez}, \citenamefont {Neupane}, \citenamefont {Alidoust},
  \citenamefont {Liu}, \citenamefont {Wang}, \citenamefont {Lee}, \citenamefont
  {Jeng}, \citenamefont {Zhang}, \citenamefont {Yuan}, \citenamefont {Jia},
  \citenamefont {Bansil}, \citenamefont {Chou}, \citenamefont {Lin},\ and\
  \citenamefont {Hasan}}]{Bian2016}%
  \BibitemOpen
  \bibfield  {author} {\bibinfo {author} {\bibfnamefont {G.}~\bibnamefont
  {Bian}}, \bibinfo {author} {\bibfnamefont {T.-R.}\ \bibnamefont {Chang}},
  \bibinfo {author} {\bibfnamefont {R.}~\bibnamefont {Sankar}}, \bibinfo
  {author} {\bibfnamefont {S.-Y.}\ \bibnamefont {Xu}}, \bibinfo {author}
  {\bibfnamefont {H.}~\bibnamefont {Zheng}}, \bibinfo {author} {\bibfnamefont
  {T.}~\bibnamefont {Neupert}}, \bibinfo {author} {\bibfnamefont {C.-K.}\
  \bibnamefont {Chiu}}, \bibinfo {author} {\bibfnamefont {S.-M.}\ \bibnamefont
  {Huang}}, \bibinfo {author} {\bibfnamefont {G.}~\bibnamefont {Chang}},
  \bibinfo {author} {\bibfnamefont {I.}~\bibnamefont {Belopolski}}, \bibinfo
  {author} {\bibfnamefont {D.~S.}\ \bibnamefont {Sanchez}}, \bibinfo {author}
  {\bibfnamefont {M.}~\bibnamefont {Neupane}}, \bibinfo {author} {\bibfnamefont
  {N.}~\bibnamefont {Alidoust}}, \bibinfo {author} {\bibfnamefont
  {C.}~\bibnamefont {Liu}}, \bibinfo {author} {\bibfnamefont {B.}~\bibnamefont
  {Wang}}, \bibinfo {author} {\bibfnamefont {C.-C.}\ \bibnamefont {Lee}},
  \bibinfo {author} {\bibfnamefont {H.-T.}\ \bibnamefont {Jeng}}, \bibinfo
  {author} {\bibfnamefont {C.}~\bibnamefont {Zhang}}, \bibinfo {author}
  {\bibfnamefont {Z.}~\bibnamefont {Yuan}}, \bibinfo {author} {\bibfnamefont
  {S.}~\bibnamefont {Jia}}, \bibinfo {author} {\bibfnamefont {A.}~\bibnamefont
  {Bansil}}, \bibinfo {author} {\bibfnamefont {F.}~\bibnamefont {Chou}},
  \bibinfo {author} {\bibfnamefont {H.}~\bibnamefont {Lin}},\ and\ \bibinfo
  {author} {\bibfnamefont {M.~Z.}\ \bibnamefont {Hasan}},\ }\bibfield  {title}
  {\bibinfo {title} {Topological nodal-line fermions in spin-orbit metal
  pbtase2},\ }\href@noop {} {\bibfield  {journal} {\bibinfo  {journal} {Nature
  Communications}\ }\textbf {\bibinfo {volume} {7}},\ \bibinfo {pages} {10556}
  (\bibinfo {year} {2016})}\BibitemShut {NoStop}%
\bibitem [{\citenamefont {Wu}\ \emph {et~al.}(2016)\citenamefont {Wu},
  \citenamefont {Wang}, \citenamefont {Mun}, \citenamefont {Johnson},
  \citenamefont {Mou}, \citenamefont {Huang}, \citenamefont {Lee},
  \citenamefont {Bud'ko}, \citenamefont {Canfield},\ and\ \citenamefont
  {Kaminski}}]{Wu2016}%
  \BibitemOpen
  \bibfield  {author} {\bibinfo {author} {\bibfnamefont {Y.}~\bibnamefont
  {Wu}}, \bibinfo {author} {\bibfnamefont {L.-L.}\ \bibnamefont {Wang}},
  \bibinfo {author} {\bibfnamefont {E.}~\bibnamefont {Mun}}, \bibinfo {author}
  {\bibfnamefont {D.~D.}\ \bibnamefont {Johnson}}, \bibinfo {author}
  {\bibfnamefont {D.}~\bibnamefont {Mou}}, \bibinfo {author} {\bibfnamefont
  {L.}~\bibnamefont {Huang}}, \bibinfo {author} {\bibfnamefont
  {Y.}~\bibnamefont {Lee}}, \bibinfo {author} {\bibfnamefont {S.~L.}\
  \bibnamefont {Bud'ko}}, \bibinfo {author} {\bibfnamefont {P.~C.}\
  \bibnamefont {Canfield}},\ and\ \bibinfo {author} {\bibfnamefont
  {A.}~\bibnamefont {Kaminski}},\ }\bibfield  {title} {\bibinfo {title} {Dirac
  node arcs in ptsn4},\ }\href@noop {} {\bibfield  {journal} {\bibinfo
  {journal} {Nature Physics}\ }\textbf {\bibinfo {volume} {12}},\ \bibinfo
  {pages} {667} (\bibinfo {year} {2016})}\BibitemShut {NoStop}%
\bibitem [{\citenamefont {Schoop}\ \emph {et~al.}(2016)\citenamefont {Schoop},
  \citenamefont {Ali}, \citenamefont {Straßer}, \citenamefont {Topp},
  \citenamefont {Varykhalov}, \citenamefont {Marchenko}, \citenamefont
  {Duppel}, \citenamefont {Parkin}, \citenamefont {Lotsch},\ and\ \citenamefont
  {Ast}}]{Schoop2016}%
  \BibitemOpen
  \bibfield  {author} {\bibinfo {author} {\bibfnamefont {L.~M.}\ \bibnamefont
  {Schoop}}, \bibinfo {author} {\bibfnamefont {M.~N.}\ \bibnamefont {Ali}},
  \bibinfo {author} {\bibfnamefont {C.}~\bibnamefont {Straßer}}, \bibinfo
  {author} {\bibfnamefont {A.}~\bibnamefont {Topp}}, \bibinfo {author}
  {\bibfnamefont {A.}~\bibnamefont {Varykhalov}}, \bibinfo {author}
  {\bibfnamefont {D.}~\bibnamefont {Marchenko}}, \bibinfo {author}
  {\bibfnamefont {V.}~\bibnamefont {Duppel}}, \bibinfo {author} {\bibfnamefont
  {S.~S.~P.}\ \bibnamefont {Parkin}}, \bibinfo {author} {\bibfnamefont {B.~V.}\
  \bibnamefont {Lotsch}},\ and\ \bibinfo {author} {\bibfnamefont {C.~R.}\
  \bibnamefont {Ast}},\ }\bibfield  {title} {\bibinfo {title} {Dirac cone
  protected by non-symmorphic symmetry and three-dimensional dirac line node in
  zrsis},\ }\href@noop {} {\bibfield  {journal} {\bibinfo  {journal} {Nature
  Communications}\ }\textbf {\bibinfo {volume} {7}},\ \bibinfo {pages} {11696}
  (\bibinfo {year} {2016})}\BibitemShut {NoStop}%
\bibitem [{\citenamefont {Neupane}\ \emph {et~al.}(2016)\citenamefont
  {Neupane}, \citenamefont {Belopolski}, \citenamefont {Hosen}, \citenamefont
  {Sanchez}, \citenamefont {Sankar}, \citenamefont {Szlawska}, \citenamefont
  {Xu}, \citenamefont {Dimitri}, \citenamefont {Dhakal}, \citenamefont
  {Maldonado}, \citenamefont {Oppeneer}, \citenamefont {Kaczorowski},
  \citenamefont {Chou}, \citenamefont {Hasan},\ and\ \citenamefont
  {Durakiewicz}}]{Neupane2016}%
  \BibitemOpen
  \bibfield  {author} {\bibinfo {author} {\bibfnamefont {M.}~\bibnamefont
  {Neupane}}, \bibinfo {author} {\bibfnamefont {I.}~\bibnamefont {Belopolski}},
  \bibinfo {author} {\bibfnamefont {M.~M.}\ \bibnamefont {Hosen}}, \bibinfo
  {author} {\bibfnamefont {D.~S.}\ \bibnamefont {Sanchez}}, \bibinfo {author}
  {\bibfnamefont {R.}~\bibnamefont {Sankar}}, \bibinfo {author} {\bibfnamefont
  {M.}~\bibnamefont {Szlawska}}, \bibinfo {author} {\bibfnamefont {S.-Y.}\
  \bibnamefont {Xu}}, \bibinfo {author} {\bibfnamefont {K.}~\bibnamefont
  {Dimitri}}, \bibinfo {author} {\bibfnamefont {N.}~\bibnamefont {Dhakal}},
  \bibinfo {author} {\bibfnamefont {P.}~\bibnamefont {Maldonado}}, \bibinfo
  {author} {\bibfnamefont {P.~M.}\ \bibnamefont {Oppeneer}}, \bibinfo {author}
  {\bibfnamefont {D.}~\bibnamefont {Kaczorowski}}, \bibinfo {author}
  {\bibfnamefont {F.}~\bibnamefont {Chou}}, \bibinfo {author} {\bibfnamefont
  {M.~Z.}\ \bibnamefont {Hasan}},\ and\ \bibinfo {author} {\bibfnamefont
  {T.}~\bibnamefont {Durakiewicz}},\ }\bibfield  {title} {\bibinfo {title}
  {Observation of topological nodal fermion semimetal phase in zrsis},\
  }\href@noop {} {\bibfield  {journal} {\bibinfo  {journal} {Phys. Rev. B}\
  }\textbf {\bibinfo {volume} {93}},\ \bibinfo {pages} {201104(R)} (\bibinfo
  {year} {2016})}\BibitemShut {NoStop}%
\bibitem [{\citenamefont {Yang}\ \emph {et~al.}(2018)\citenamefont {Yang},
  \citenamefont {Yang}, \citenamefont {Derunova}, \citenamefont {Parkin},
  \citenamefont {Yan},\ and\ \citenamefont {Ali}}]{Yang2018}%
  \BibitemOpen
  \bibfield  {author} {\bibinfo {author} {\bibfnamefont {S.-Y.}\ \bibnamefont
  {Yang}}, \bibinfo {author} {\bibfnamefont {H.}~\bibnamefont {Yang}}, \bibinfo
  {author} {\bibfnamefont {E.}~\bibnamefont {Derunova}}, \bibinfo {author}
  {\bibfnamefont {S.~S.~P.}\ \bibnamefont {Parkin}}, \bibinfo {author}
  {\bibfnamefont {B.}~\bibnamefont {Yan}},\ and\ \bibinfo {author}
  {\bibfnamefont {M.~N.}\ \bibnamefont {Ali}},\ }\bibfield  {title} {\bibinfo
  {title} {Symmetry demanded topological nodal-line materials},\ }\href@noop {}
  {\bibfield  {journal} {\bibinfo  {journal} {Advances in Physics: X}\ }\textbf
  {\bibinfo {volume} {3}},\ \bibinfo {pages} {1414631} (\bibinfo {year}
  {2018})}\BibitemShut {NoStop}%
\bibitem [{\citenamefont {Klemenz}\ \emph {et~al.}(2019)\citenamefont
  {Klemenz}, \citenamefont {Lei},\ and\ \citenamefont {Schoop}}]{Klemenz2019}%
  \BibitemOpen
  \bibfield  {author} {\bibinfo {author} {\bibfnamefont {S.}~\bibnamefont
  {Klemenz}}, \bibinfo {author} {\bibfnamefont {S.}~\bibnamefont {Lei}},\ and\
  \bibinfo {author} {\bibfnamefont {L.~M.}\ \bibnamefont {Schoop}},\ }\bibfield
   {title} {\bibinfo {title} {Topological semimetals in square-net materials},\
  }\href@noop {} {\bibfield  {journal} {\bibinfo  {journal} {Annual Review of
  Materials Research}\ }\textbf {\bibinfo {volume} {49}},\ \bibinfo {pages}
  {185} (\bibinfo {year} {2019})}\BibitemShut {NoStop}%
\bibitem [{\citenamefont {Ali}\ \emph {et~al.}(2016)\citenamefont {Ali},
  \citenamefont {Schoop}, \citenamefont {Garg}, \citenamefont {Lippmann},
  \citenamefont {Lara}, \citenamefont {Lotsch},\ and\ \citenamefont
  {Parkin}}]{Ali2016}%
  \BibitemOpen
  \bibfield  {author} {\bibinfo {author} {\bibfnamefont {M.~N.}\ \bibnamefont
  {Ali}}, \bibinfo {author} {\bibfnamefont {L.~M.}\ \bibnamefont {Schoop}},
  \bibinfo {author} {\bibfnamefont {C.}~\bibnamefont {Garg}}, \bibinfo {author}
  {\bibfnamefont {J.~M.}\ \bibnamefont {Lippmann}}, \bibinfo {author}
  {\bibfnamefont {E.}~\bibnamefont {Lara}}, \bibinfo {author} {\bibfnamefont
  {B.}~\bibnamefont {Lotsch}},\ and\ \bibinfo {author} {\bibfnamefont
  {S.~S.~P.}\ \bibnamefont {Parkin}},\ }\bibfield  {title} {\bibinfo {title}
  {Butterfly magnetoresistance, quasi-2d dirac fermi surface and topological
  phase transition in zrsis},\ }\href@noop {} {\bibfield  {journal} {\bibinfo
  {journal} {Science Advances}\ }\textbf {\bibinfo {volume} {2}},\ \bibinfo
  {pages} {e1601742} (\bibinfo {year} {2016})}\BibitemShut {NoStop}%
\bibitem [{\citenamefont {Gatti}\ \emph {et~al.}(2020)\citenamefont {Gatti},
  \citenamefont {Crepaldi}, \citenamefont {Puppin}, \citenamefont
  {Tancogne-Dejean}, \citenamefont {Xian}, \citenamefont {De~Giovannini},
  \citenamefont {Roth}, \citenamefont {Polishchuk}, \citenamefont {Bugnon},
  \citenamefont {Magrez}, \citenamefont {Berger}, \citenamefont {Frassetto},
  \citenamefont {Poletto}, \citenamefont {Moreschini}, \citenamefont {Moser},
  \citenamefont {Bostwick}, \citenamefont {Rotenberg}, \citenamefont {Rubio},
  \citenamefont {Chergui},\ and\ \citenamefont {Grioni}}]{Gatti2020}%
  \BibitemOpen
  \bibfield  {author} {\bibinfo {author} {\bibfnamefont {G.}~\bibnamefont
  {Gatti}}, \bibinfo {author} {\bibfnamefont {A.}~\bibnamefont {Crepaldi}},
  \bibinfo {author} {\bibfnamefont {M.}~\bibnamefont {Puppin}}, \bibinfo
  {author} {\bibfnamefont {N.}~\bibnamefont {Tancogne-Dejean}}, \bibinfo
  {author} {\bibfnamefont {L.}~\bibnamefont {Xian}}, \bibinfo {author}
  {\bibfnamefont {U.}~\bibnamefont {De~Giovannini}}, \bibinfo {author}
  {\bibfnamefont {S.}~\bibnamefont {Roth}}, \bibinfo {author} {\bibfnamefont
  {S.}~\bibnamefont {Polishchuk}}, \bibinfo {author} {\bibfnamefont
  {P.}~\bibnamefont {Bugnon}}, \bibinfo {author} {\bibfnamefont
  {A.}~\bibnamefont {Magrez}}, \bibinfo {author} {\bibfnamefont
  {H.}~\bibnamefont {Berger}}, \bibinfo {author} {\bibfnamefont
  {F.}~\bibnamefont {Frassetto}}, \bibinfo {author} {\bibfnamefont
  {L.}~\bibnamefont {Poletto}}, \bibinfo {author} {\bibfnamefont
  {L.}~\bibnamefont {Moreschini}}, \bibinfo {author} {\bibfnamefont
  {S.}~\bibnamefont {Moser}}, \bibinfo {author} {\bibfnamefont
  {A.}~\bibnamefont {Bostwick}}, \bibinfo {author} {\bibfnamefont
  {E.}~\bibnamefont {Rotenberg}}, \bibinfo {author} {\bibfnamefont
  {A.}~\bibnamefont {Rubio}}, \bibinfo {author} {\bibfnamefont
  {M.}~\bibnamefont {Chergui}},\ and\ \bibinfo {author} {\bibfnamefont
  {M.}~\bibnamefont {Grioni}},\ }\bibfield  {title} {\bibinfo {title}
  {Light-induced renormalization of the dirac quasiparticles in the nodal-line
  semimetal zrsise},\ }\href@noop {} {\bibfield  {journal} {\bibinfo  {journal}
  {Phys. Rev. Lett.}\ }\textbf {\bibinfo {volume} {125}},\ \bibinfo {pages}
  {076401} (\bibinfo {year} {2020})}\BibitemShut {NoStop}%
\bibitem [{\citenamefont {Shao}\ \emph {et~al.}(2020)\citenamefont {Shao},
  \citenamefont {Rudenko}, \citenamefont {Hu}, \citenamefont {Sun},
  \citenamefont {Zhu}, \citenamefont {Moon}, \citenamefont {Millis},
  \citenamefont {Yuan}, \citenamefont {Lichtenstein}, \citenamefont {Smirnov},
  \citenamefont {Mao}, \citenamefont {Katsnelson},\ and\ \citenamefont
  {Basov}}]{Shao2020}%
  \BibitemOpen
  \bibfield  {author} {\bibinfo {author} {\bibfnamefont {Y.}~\bibnamefont
  {Shao}}, \bibinfo {author} {\bibfnamefont {A.~N.}\ \bibnamefont {Rudenko}},
  \bibinfo {author} {\bibfnamefont {J.}~\bibnamefont {Hu}}, \bibinfo {author}
  {\bibfnamefont {Z.}~\bibnamefont {Sun}}, \bibinfo {author} {\bibfnamefont
  {Y.}~\bibnamefont {Zhu}}, \bibinfo {author} {\bibfnamefont {S.}~\bibnamefont
  {Moon}}, \bibinfo {author} {\bibfnamefont {A.~J.}\ \bibnamefont {Millis}},
  \bibinfo {author} {\bibfnamefont {S.}~\bibnamefont {Yuan}}, \bibinfo {author}
  {\bibfnamefont {A.~I.}\ \bibnamefont {Lichtenstein}}, \bibinfo {author}
  {\bibfnamefont {D.}~\bibnamefont {Smirnov}}, \bibinfo {author} {\bibfnamefont
  {Z.~Q.}\ \bibnamefont {Mao}}, \bibinfo {author} {\bibfnamefont {M.~I.}\
  \bibnamefont {Katsnelson}},\ and\ \bibinfo {author} {\bibfnamefont {D.~N.}\
  \bibnamefont {Basov}},\ }\bibfield  {title} {\bibinfo {title} {Electronic
  correlations in nodal-line semimetals},\ }\href@noop {} {\bibfield  {journal}
  {\bibinfo  {journal} {Nat. Phys.}\ }\textbf {\bibinfo {volume} {16}},\
  \bibinfo {pages} {636–641} (\bibinfo {year} {2020})}\BibitemShut {NoStop}%
\bibitem [{\citenamefont {Ramamurthy}\ and\ \citenamefont
  {Hughes}(2017)}]{Ramamurthy2017}%
  \BibitemOpen
  \bibfield  {author} {\bibinfo {author} {\bibfnamefont {S.~T.}\ \bibnamefont
  {Ramamurthy}}\ and\ \bibinfo {author} {\bibfnamefont {T.~L.}\ \bibnamefont
  {Hughes}},\ }\bibfield  {title} {\bibinfo {title} {Quasitopological
  electromagnetic response of line-node semimetals},\ }\href@noop {} {\bibfield
   {journal} {\bibinfo  {journal} {Phys. Rev. B.}\ }\textbf {\bibinfo {volume}
  {95}},\ \bibinfo {pages} {075138} (\bibinfo {year} {2017})}\BibitemShut
  {NoStop}%
\bibitem [{\citenamefont {Rudenko}\ \emph {et~al.}(2018)\citenamefont
  {Rudenko}, \citenamefont {Stepanov}, \citenamefont {Lichtenstein},\ and\
  \citenamefont {Katsnelson}}]{Rudenko2018}%
  \BibitemOpen
  \bibfield  {author} {\bibinfo {author} {\bibfnamefont {A.~N.}\ \bibnamefont
  {Rudenko}}, \bibinfo {author} {\bibfnamefont {E.~A.}\ \bibnamefont
  {Stepanov}}, \bibinfo {author} {\bibfnamefont {A.~I.}\ \bibnamefont
  {Lichtenstein}},\ and\ \bibinfo {author} {\bibfnamefont {M.~I.}\ \bibnamefont
  {Katsnelson}},\ }\bibfield  {title} {\bibinfo {title} {Excitonic instability
  and pseudogap formation in nodal line semimetal zrsis},\ }\href@noop {}
  {\bibfield  {journal} {\bibinfo  {journal} {Phys. Rev. Lett.}\ }\textbf
  {\bibinfo {volume} {120}},\ \bibinfo {pages} {216401} (\bibinfo {year}
  {2018})}\BibitemShut {NoStop}%
\bibitem [{\citenamefont {Huh}\ \emph {et~al.}(2016)\citenamefont {Huh},
  \citenamefont {Moon},\ and\ \citenamefont {Kim}}]{Huh2016}%
  \BibitemOpen
  \bibfield  {author} {\bibinfo {author} {\bibfnamefont {Y.}~\bibnamefont
  {Huh}}, \bibinfo {author} {\bibfnamefont {E.-G.}\ \bibnamefont {Moon}},\ and\
  \bibinfo {author} {\bibfnamefont {Y.~B.}\ \bibnamefont {Kim}},\ }\bibfield
  {title} {\bibinfo {title} {Long-range coulomb interaction in nodal-ring
  semimetals},\ }\href@noop {} {\bibfield  {journal} {\bibinfo  {journal}
  {Phys. Rev. B}\ }\textbf {\bibinfo {volume} {93}},\ \bibinfo {pages} {035138}
  (\bibinfo {year} {2016})}\BibitemShut {NoStop}%
\bibitem [{\citenamefont {Syzranov}\ and\ \citenamefont
  {Skinner}(2017)}]{Syzranov2017}%
  \BibitemOpen
  \bibfield  {author} {\bibinfo {author} {\bibfnamefont {S.~V.}\ \bibnamefont
  {Syzranov}}\ and\ \bibinfo {author} {\bibfnamefont {B.}~\bibnamefont
  {Skinner}},\ }\bibfield  {title} {\bibinfo {title} {Electron transport in
  nodal-line semimetals},\ }\href@noop {} {\bibfield  {journal} {\bibinfo
  {journal} {Phys. Rev. B}\ }\textbf {\bibinfo {volume} {96}},\ \bibinfo
  {pages} {161105(R)} (\bibinfo {year} {2017})}\BibitemShut {NoStop}%
\bibitem [{\citenamefont {Narozhny}\ and\ \citenamefont
  {Gornyi}(2021)}]{Narozhny2021}%
  \BibitemOpen
  \bibfield  {author} {\bibinfo {author} {\bibfnamefont {B.~N.}\ \bibnamefont
  {Narozhny}}\ and\ \bibinfo {author} {\bibfnamefont {I.~V.}\ \bibnamefont
  {Gornyi}},\ }\bibfield  {title} {\bibinfo {title} {Hydrodynamic approach to
  electronic transport in graphene: Energy relaxation},\ }\href@noop {}
  {\bibfield  {journal} {\bibinfo  {journal} {Frontiers in Physics}\ }\textbf
  {\bibinfo {volume} {9}},\ \bibinfo {pages} {108} (\bibinfo {year}
  {2021})}\BibitemShut {NoStop}%
\bibitem [{\citenamefont {Lucas}\ and\ \citenamefont {Fong}(2018)}]{Lucas2018}%
  \BibitemOpen
  \bibfield  {author} {\bibinfo {author} {\bibfnamefont {A.}~\bibnamefont
  {Lucas}}\ and\ \bibinfo {author} {\bibfnamefont {K.~C.}\ \bibnamefont
  {Fong}},\ }\bibfield  {title} {\bibinfo {title} {Hydrodynamics of electrons
  in graphene},\ }\href@noop {} {\bibfield  {journal} {\bibinfo  {journal}
  {Journal of Physics: Condensed Matter}\ }\textbf {\bibinfo {volume} {30}},\
  \bibinfo {pages} {053001} (\bibinfo {year} {2018})}\BibitemShut {NoStop}%
\bibitem [{\citenamefont {Rudenko}\ and\ \citenamefont
  {Yuan}(2020)}]{Rudenko2020}%
  \BibitemOpen
  \bibfield  {author} {\bibinfo {author} {\bibfnamefont {A.~N.}\ \bibnamefont
  {Rudenko}}\ and\ \bibinfo {author} {\bibfnamefont {S.}~\bibnamefont {Yuan}},\
  }\bibfield  {title} {\bibinfo {title} {Electron-phonon interaction and
  zero-field charge carrier transport in the nodal-line semimetal zrsis},\
  }\href@noop {} {\bibfield  {journal} {\bibinfo  {journal} {Phys. Rev. B}\
  }\textbf {\bibinfo {volume} {101}},\ \bibinfo {pages} {115127} (\bibinfo
  {year} {2020})}\BibitemShut {NoStop}%
\bibitem [{\citenamefont {Chang}\ \emph {et~al.}(2016)\citenamefont {Chang},
  \citenamefont {Chen}, \citenamefont {Bian}, \citenamefont {Huang},
  \citenamefont {Zheng}, \citenamefont {Neupert}, \citenamefont {Sankar},
  \citenamefont {Xu}, \citenamefont {Belopolski}, \citenamefont {Chang},
  \citenamefont {Wang}, \citenamefont {Chou}, \citenamefont {Bansil},
  \citenamefont {Jeng}, \citenamefont {Lin},\ and\ \citenamefont
  {Hasan}}]{Chang2016}%
  \BibitemOpen
  \bibfield  {author} {\bibinfo {author} {\bibfnamefont {T.-R.}\ \bibnamefont
  {Chang}}, \bibinfo {author} {\bibfnamefont {P.-J.}\ \bibnamefont {Chen}},
  \bibinfo {author} {\bibfnamefont {G.}~\bibnamefont {Bian}}, \bibinfo {author}
  {\bibfnamefont {S.-M.}\ \bibnamefont {Huang}}, \bibinfo {author}
  {\bibfnamefont {H.}~\bibnamefont {Zheng}}, \bibinfo {author} {\bibfnamefont
  {T.}~\bibnamefont {Neupert}}, \bibinfo {author} {\bibfnamefont
  {R.}~\bibnamefont {Sankar}}, \bibinfo {author} {\bibfnamefont {S.-Y.}\
  \bibnamefont {Xu}}, \bibinfo {author} {\bibfnamefont {I.}~\bibnamefont
  {Belopolski}}, \bibinfo {author} {\bibfnamefont {G.}~\bibnamefont {Chang}},
  \bibinfo {author} {\bibfnamefont {B.~K.}\ \bibnamefont {Wang}}, \bibinfo
  {author} {\bibfnamefont {F.}~\bibnamefont {Chou}}, \bibinfo {author}
  {\bibfnamefont {A.}~\bibnamefont {Bansil}}, \bibinfo {author} {\bibfnamefont
  {H.-T.}\ \bibnamefont {Jeng}}, \bibinfo {author} {\bibfnamefont
  {H.}~\bibnamefont {Lin}},\ and\ \bibinfo {author} {\bibfnamefont {M.~Z.}\
  \bibnamefont {Hasan}},\ }\bibfield  {title} {\bibinfo {title} {Topological
  dirac surface states and superconducting pairing correlations in
  ${\mathrm{pbtase}}_{2}$},\ }\href@noop {} {\bibfield  {journal} {\bibinfo
  {journal} {Phys. Rev. B}\ }\textbf {\bibinfo {volume} {93}},\ \bibinfo
  {pages} {245130} (\bibinfo {year} {2016})}\BibitemShut {NoStop}%
\end{thebibliography}

%

\end{document}